\documentclass[11pt,a4paper]{article}
\usepackage{amsfonts}

\usepackage{graphicx}
\usepackage{amsmath}
\usepackage{a4wide}

\input{tcilatex}

\begin{document}

\title{\textbf{General Framework for the Behaviour of Continuously Observed Open
Quantum Systems}}
\author{\textbf{Ole E. Barndorff-Nielsen}\thanks{%
E-mail: oebn@imf.au.dk} \\
MaPhySto\thanks{%
MaPhySto - Centre for Mathematical Physics and Stochastics, funded by the
Danish National Research Foundation}, University of Aarhus, DK-8000 Aarhus,
Denmark \and \textbf{Elena R. Loubenets}\thanks{%
E-mail: elena@imf.au.dk, erl@erl.msk.ru} \\
Moscow State Institute of Electronics and Mathematics, Technical University, 
\\
Trekhsvyatitelskii Per. 3/12, Moscow 109028, Russia \\
\\
Published in: \textit{J.Physics A: Math. Gen.}\textbf{\ }v.35 (2002), N3,
565-588 \\
http://stacks.iop.org/JPhysA}
\maketitle

\begin{abstract}
We develop the general quantum stochastic approach to the description of
quantum measurements continuous in time. The framework, that we introduce,
encompasses the various particular models for continuous-time measurements
considered previously in the physical and the mathematical literature.
\end{abstract}

\tableofcontents

\section{Introduction}

In recent years, the problem of how to describe the behaviour of a quantum
system continuously observed in time has been the subject of intensive
investigations in the physical and the mathematical literature. This type of
behaviour is generally not reversible in time and, hence, in particular, can
not be described by the Schr\"{o}dinger equation whose solutions are
reversible. The present strong interest in this fundamental problem is to a
large extent caused by the rapid development of experimental techniques,
where experiments involving continuous-time (i.e. continuous in time)
`monitoring' of a quantum system have become possible [19-21,31,41].

The present paper develops, in the context of continuous-time monitoring,
the general approach to the description of quantum measurements, formulated
in [36,37]. The general framework, that we introduce, encompasses the
various particular models for continuous-time measurements, considered
previously in the physical and the mathematical literature. This concerns:

\begin{itemize}
\item  The Markovian models in the mathematical physics literature [1-10,
27,30], formulated either in terms of stochastic differential equations or
in terms of semigroups of probability operators or in terms of the
instrumental processes with independent increments. In fact, the stochastic
equations used in all these models generalize the quantum filtering equation
([7-10] and references therein) which was derived in the quantum stochastic
calculus modelling framework. This modelling framework satisfies the
principles of nondemolition observation, introduced in [7-10]. It was,
however, shown in [35] that the case of continuous-time indirect
nondemolition measurement can be considered in the more general quantum
theory setting, which is not based on the use of the essentially Markovian
measurement model of quantum stochastic calculus.

\item  The models of continuous-time observation in the physical literature
[11,13,14,17,22-25,34,38,39,44,47-51], including those in quantum optics.
The derivation of stochastic equations in all these models is based mostly
on unraveling of the Master equation of Lindblad type [33] (cf., for
example, [39, 50,51]) or on the phenomenological introduction of
non-Markovian quantum trajectories [47-49].
\end{itemize}

As a prerequisite for our results in the main part of the paper, we review
in Section 2 the main concepts of the operational approach (subsection 2.1)
and the main ideas of the quantum stochastic approach (QSA) (subsection 2.2)
to the description of quantum measurements.

Sections 3-8 then develop the formalism for the description of
continuous-time measurements from the general viewpoint of the QSA,
formulated in [36,37]. The general scene is set in Section 3. Section 4
introduces the notion of a posterior\ pure state trajectory and gives its
probabilistic treatment. The special case of Markov evolution is treated in
Section 5. Section 6 establishes the notion of a measuring model of
continuous-time direct quantum measurement and Sections 7, 8 address the
questions of continuous-time nondemolition measurement. The final Section 9
consists of concluding remarks.

\section{Basic representations of quantum measurements}

This Section reviews both the operational approach and the quantum
stochastic approach to the description of quantum measurements.

By a quantum measurement we mean a physical experiment upon a quantum system
which, resulting in the observation in the classical world of an outcome
that to some degree characterizes the quantum system, may cause a change in
the state of the quantum system, but not the quantum system's destruction.

We distinguish between\textit{\ }\emph{direct} and \emph{indirect} quantum
measurements.

A direct quantum measurement corresponds to a measurement situation where we
have to describe the direct interaction between the measuring device and the
observed quantum system, while in case of an indirect measurement, a direct
measurement is made of some other quantum system, entangled with the one
considered.

The term `generalized measurement', as usual, corresponds to the measurement
situation with outcomes of the most general nature possible under a quantum
measurement.

Let a quantum system $S$, described in terms of a complex separable Hilbert
space $\mathcal{H}$, interact with another system (quantum or classical).
The interaction, changing the initial state $\rho _{0}$ of $S$ into a
certain new state, leaves some imprint in the classical world, the imprint
being described as a point $\omega $ in some standard Borel measure space 
\footnote{%
A standard Borel space is a measurable space that is isomorphic to the real
unit interval. In particular, any Polish space is standard Borel.}$(\Omega ,%
\mathcal{F})$. Denote by $\mathcal{B(H)}$ the Banach space of all bounded
linear operators on $\mathcal{H}$.

Consider first the most general scheme of the \textit{complete statistical
description} of any generalized quantum measurement. This kind of
description implies the knowledge of the probability distribution of
different outcomes of a measurement and a statistical description of the
state change of the quantum system under this measurement.

We introduce the following notation.

Let $\ \pi (B;\rho _{0})=\Pr $ob$\{\omega \in B;\rho _{0}\}$\ be the
probability that the imprint $\omega $ in the classical world belongs to a
subset $B\in \mathcal{F}$.

Let $\mathrm{Ex}\{Z;\rho _{0}|B\}$\ be the conditional expectation of any
von Neumann observable$\ Z=Z^{\ast },Z\in \mathcal{B}(\mathcal{H}),$ at the
instant immediately after the measurement and conditioned on the outcome $%
\omega \in B.$\ 

The posterior state (or posterior density operator)$\ \rho _{out}(B;\rho
_{0})$, of the quantum system conditioned by the imprint $B\ $in the
classical world, is defined indirectly as the solution to 
\begin{equation}
\mathrm{Ex}\{Z;\rho _{0}|B\}=tr\{\rho _{out}(B;\rho _{0})Z\},  \label{2.1}
\end{equation}
(for arbitrary $Z$ and $B$) and constitutes the statistical description of
the state change of the quantum system under a measurement when only the
event that $\omega $\ belongs to $B$ has been recorded (cf.[42,43,2,36,37]).

The\textit{\ }unconditional posterior state $\rho _{out}(\Omega ;\rho _{0})$
of the quantum system corresponds to the situation where the imprint $\omega 
$ in the classical world is ignored completely.

Any posterior state $\rho _{out}(B;\rho _{0})$ can be described in terms of
a family of statistical operators $\{\rho _{out}(\omega ;\rho _{0}),\omega
\in \Omega \}$, defined $\ \pi $-almost everywhere (a.e.) on $\Omega $, and
usually referred to as the \textit{family of posterior states.}
Specifically, for all $B\in \mathcal{F}$ with $\pi (B;\rho _{0})\neq 0,$ 
\begin{equation}
\rho _{out}(B;\rho _{0})=\frac{\int_{B}\rho _{out}(\omega ;\rho _{0})\pi
(d\omega ;\rho _{0})}{\pi (B;\rho _{0})}  \label{2.2}
\end{equation}

For the unconditional posterior state $\rho _{out}(\Omega ;\rho _{0})$ the
relation (\ref{2.2}) can be considered as the usual statistical average over
the posterior states $\rho _{out}(\omega ;\rho _{0})$ with respect to the
probability distribution $\pi (d\omega ;\rho _{0})$.

For any type (direct or indirect) of a generalized quantum measurement the
operational approach [15,16,26,32,42,43,2,29] can be used for the most
general mathematical specification of all of the above-mentioned elements of
the statistical description of a measurement.

\subsection{The operational description of a generalized quantum measurement}

In the frame of the operational approach the mathematical notion of a \emph{%
quantum instrument} plays a central role.

Specifically, a mapping $\mathcal{N}(\cdot )[\cdot ]$: $\mathcal{F\times
B(H)\rightarrow B(H)}$ is called a quantum instrument\ if $\mathcal{N}(\cdot
)$ is a $\sigma $-additive measure on $(\Omega ,\mathcal{F)}$ with values $%
\mathcal{N}(B),$ $B\in \mathcal{F}$, that are normal completely positive%
\footnote{%
For the definitions of normality and complete positivity see, for instance,
[29]} bounded linear maps $\mathcal{B(H)}\rightarrow \mathcal{B(H)}$ such
that the following normalisation relation is valid: $\mathcal{N}(\Omega
)[I]=I$.

In the frame of the operational approach it is assumed\textit{\ }that 
\begin{equation}
\mathrm{Ex}\{Z;\rho _{0}|B\}=\frac{\mathrm{tr}\{\rho _{0}\mathcal{N}(B)[Z]\}%
}{\pi (B;\rho _{0})},\text{ \ \ }\forall B\in \mathcal{F}\text{.}
\label{2.3}
\end{equation}
In case $Z=I$, from (\ref{2.3}) it follows that the probability distribution 
$\pi (B;\rho _{0})$ of outcomes under a measurement is given by 
\begin{equation}
\pi (B;\rho _{0})=\mathrm{tr}\{\rho _{0}\mathcal{N}(B)[I]\},\text{ \ \ \ }%
\forall B\in \mathcal{F}\text{.}  \label{2.4}
\end{equation}

The positive $\sigma $-additive operator-valued measure $M(B)=\mathcal{N}%
(B)[I]$, satisfying the condition $M(\Omega )=I$, is called a \textit{%
probability operator-valued measure} or a POV measure, for short.

Due to (\ref{2.3}), in the frame of the operational approach the posterior
state $\rho _{out}(B;\rho _{0})$, conditioned by the outcome $\omega \in B$
and defined by the relation (\ref{2.1}), is representable as 
\begin{equation}
\rho _{out}(B;\rho _{0})=\frac{\mathcal{M}(B)[\rho _{0}]}{\pi (B;\rho _{0})},
\label{2.5}
\end{equation}
where $\mathcal{M}(B)[\cdot ]$ denotes the map dual to $\mathcal{N}(B)[\cdot
]$, which acts on the Banach space $\mathcal{T(H)}$ of trace-class operators
on $\mathcal{H}$ and is defined by 
\begin{equation}
\mathrm{tr}\{\kappa \mathcal{N}(B)[Y]\}=\mathrm{tr}\{\mathcal{M}(B)[\kappa
]Y\},  \label{2.6}
\end{equation}
for arbitrary $Y\in \mathcal{B(H)},\ \kappa \in \mathcal{T(H)}$. Since $%
\mathcal{N}(\Omega )[I]=I$, it follows from (\ref{2.6}) that $\mathrm{tr}\{%
\mathcal{M}(\Omega )[\rho ]\}=1$ for any density operator $\rho \in \mathcal{%
T(H)}$. We follow the terminology of [29] and refer to $\mathcal{M}(\cdot
)[\cdot ]$ \ as a quantum instrument associated with the quantum instrument $%
\mathcal{N}(\cdot )[\cdot ]$. Due to (\ref{2.5}) we also have 
\begin{equation}
\pi (B;\rho _{0})=\mathrm{tr}\{\mathcal{M}(B)[\rho _{0}]\},\text{ \ \ }%
\forall B\in \mathcal{F}.  \label{2.7}
\end{equation}

For any initial state $\rho _{0}$ of a quantum system the family of
posterior states $\{\rho _{out}(\omega ,\rho _{0}),\omega \in \Omega \}$
always exists [42,43,2] and is defined uniquely, $\pi $-almost everywhere,
by the relation: 
\begin{equation}
\int_{B}\mathrm{tr}\{\rho _{out}(\omega ;\rho _{0})Y\}\pi (d\omega ,\rho
_{0})=\mathrm{tr}\{\rho _{0}\mathcal{N}(B)[Y]\},  \label{2.8}
\end{equation}
for all $Y\in \mathcal{B(H)},\forall B\in \mathcal{F}$. From (\ref{2.2}) and
(\ref{2.5}) we have, in particular, 
\begin{equation}
\rho _{out}(\omega ;\rho _{0})=\frac{d\mathcal{M}(\cdot )[\rho _{0}]}{d\pi
(\cdot ;\rho _{0})},  \label{2.9}
\end{equation}
that is, the posterior state $\rho _{out}(\omega ;\rho _{0})$ is a density
of the measure $\mathcal{M}(\cdot )[\rho _{0}]$\ with respect to the
probability scalar measure $\pi (\cdot ;\rho _{0})$. Further, from (\ref{2.5}%
) it follows that the unconditional posterior state is given by 
\begin{equation}
\rho _{out}(\Omega ;\rho _{0})=\mathcal{M}(\Omega )[\rho _{0}].  \label{2.10}
\end{equation}

\bigskip

It was proved in [36] that for any quantum instrument there exist:

\begin{itemize}
\item  a positive finite scalar measure $\nu (\cdot )$ on $(\Omega ,\mathcal{%
F)}$;

\item  positive integer $N_{0}\leq \infty ;$

\item  a dimension function $N(\cdot )$, defined $\nu $-almost everywhere on 
$\Omega ,$ with values being positive integers $N(\omega )\leq \infty ;$

\item  positive numbers $\alpha _{i},$ summing up to one $%
\sum_{i=1}^{N_{0}}\alpha _{i}=1;$

\item  a family $\{W_{in}:i=1,...,N_{0};n=1,...,l\}$ (with $l$ being equal
to $\nu $-$\sup \{N(\omega ),\omega \in \Omega \}$) of bounded linear
operators $W_{in}:\mathcal{H\rightarrow L}_{2}(\Omega ,\nu ;\mathcal{H}$),
satisfying for $\forall f,g\in \mathcal{H}$ the orthonormality relation 
\begin{equation}
\int_{\Omega }\sum_{n=1}^{N(\omega )}\langle (W_{jn}f)(\omega
),(W_{in}g)(\omega )\rangle \nu (d\omega )=\langle f,g\rangle \delta _{ji},
\label{2.11}
\end{equation}
such that for $\forall B\in \mathcal{F},$ $\forall Y\in \mathcal{B(H)}$ and $%
\forall f,g\in \mathcal{H}$ the following integral representation for a
quantum instrument is valid: 
\begin{equation}
\langle f,\mathcal{N(}B)[Y]g\rangle =\sum_{i=1}^{N_{0}}\alpha
_{i}\int_{B}\sum_{n=1}^{N(\omega )}\langle (W_{in}f)(\omega
),Y(W_{in}g)(\omega )\rangle \nu (d\omega ).  \label{2.12}
\end{equation}
The integral representation (\ref{2.12}) is, in general, different and more
detailed than the representations available in the mathematical and physical
literature [28,45,51]. The latter are similar to the Stinespring-Kraus
representation for a completely positive map on $\mathcal{B}(\mathcal{H)}$
(cf., for example, [29]). The most essential difference is due to the
orthonormality relation (\ref{2.11}), which is not present in the
Stinespring-Kraus like representations of a quantum instrument
[5,28,29,45,51]. Moreover, since the two different types of indexes $i,n$\
enter the orthonormality relation (\ref{2.11}) in quite different manner,
the double indexing\ in (\ref{2.12}) can not, in general, be presented as a
single one without loss of the natural structure of an orthonormality
relation (see [36] for further discussion).
\end{itemize}

We would like to underline here that having the elements of one integral
representation of an instrument one can construct, due to the definite
transformation rule (see [36]), a plenitude of other integral
representations of the same instrument with different families of operators $%
\{W_{in}^{\prime }\}$ and different scalar measures $\nu ^{\prime }$, the
latter being, however, of the same type: $[\nu ^{\prime }]=[\nu ].$

The operational approach, while essential for the formalization of the
statistical description of any generalized quantum measurement, does not, in
general, specify a possible random behaviour of the quantum system under a
single measurement. In other words, the operational approach, in general,
does not give the possibility to include into consideration the description
of the stochastic, irreversible in time behaviour of a quantum system under
a single measurement\textit{,} depending on an outcome $\omega $ in the
classical world. The description of such stochastic behaviour of a quantum
system means the specification of a probabilistic transition law governing
the change from the initial state of the quantum system to a final one under
a single quantum measurement. We refer to this kind of description of a
quantum measurement as a \textit{complete stochastic description.}

The operational approach also does not distinguish between direct and
indirect measurements.

In this connection we would like to underline that in quantum theory any
physically based problem must be formulated in unitarily equivalent terms%
\textit{\ }and the results of its consideration must not be dependent either
on the choice of a special representation picture (Schr\"{o}dinger,
Heisenberg or interaction) or on the choice of basis in the Hilbert space.
Moreover, in general, the description of any direct quantum measurement can
not be simply reduced to the quantum theory description of a measuring
process, as it is now usually considered in the mathematical and physical
literature. For this kind of measurement situation we can not specify
definitely either the interaction, or the quantum state of a measuring
device environment, or describe a measuring device in quantum theory terms
alone. In fact, under such a scheme the description of a direct quantum
measurement is simply transferred to the description of a direct measurement
of some observable of an environment of a measuring device.

We recall that for the case of discrete outcomes the original von Neumann
approach [40] describes \textit{specifically a direct quantum measurement }%
and gives both a complete statistical description and a complete stochastic
description of this measurement. Specifically, if the initial state $\rho
_{0}$ of a quantum system is pure, that is, $\rho _{0}=|\psi _{0}\rangle
\langle \psi _{0}|$,\ and if under a single measurement the outcome $\lambda
_{j}$ is observed, then in the frame of the von Neumann approach the quantum
system `jumps' with \textit{certainty }to the posterior pure state 
\begin{equation}
\frac{P_{j}|\psi _{0}\rangle \langle \psi _{0}|P_{j}}{||P_{j}\psi _{0}||},
\label{2.13}
\end{equation}
where $P_{j}$ is the projection, corresponding to the observed eigenvalue $%
\lambda _{j}$ of the observable $Z=\sum_{j}\lambda _{j}P_{j}$.\ The
probability $\mu _{j\text{ }}$ of the outcome $\lambda _{j}$ is given by 
\begin{equation}
\mu _{j\text{ }}=||P_{j}\psi _{0}||^{2}.  \label{2.14}
\end{equation}

\bigskip

An approach, giving both - a complete statistical and a complete stochastic
description of a \textit{direct }quantum measurement with outcomes of the
most general possible nature was introduced in [36,37]. This approach is
called \emph{quantum stochastic} and refer to it as QSA.

\subsection{Quantum stochastic approach}

It was shown in [36] that any generalized direct quantum measurement (cf.
Subsection 2.1 above) can be described in terms of certain scalar measures
on a standard Borel space $(\Omega ,\mathcal{F})$ and associated stochastic
evolution operators, describing the stochastic evolution of the quantum
system in the Hilbert space $\mathcal{H}$ conditioned by the observed
outcome $\omega $. We refer to the collection of these quantities as a \emph{%
quantum stochastic representation}\textit{, }or QSR, of a generalized direct
quantum measurement. For simplicity, we consider below only quantum
stochastic representations for which the quantum stochastic evolution
operators are bounded.

From the point of view of the operational approach, the QSA specifies, in
particular, the type of a quantum instrument, corresponding to the
description of a generalized direct quantum measurement.

In particular, it was shown in [36] that any \emph{generalized direct
quantum measurement} can be interpreted to correspond to an \emph{invariant
class of unitarily equivalent measuring processes} (statistical
realizations). For an invariant class of measuring processes the elements of
the integral representation (\ref{2.12}) of the corresponding instrument are
the same for all measuring processes from this class and are given only
through the \textit{unitary invariants} of the measuring process. The
special form of this integral representation of an instrument, corresponding
to the invariant class, is called \textit{quantum stochastic}.

\textit{According to the QSA, to every generalized direct quantum
measurement there exists a} \textit{unique quantum stochastic representation
of a measurement, giving a complete statistical and stochastic description
of this measurement, in a precisely specified sense. }

Specifically, by a \emph{quantum stochastic representation} (QSR), we mean a
collection 
\begin{equation}
Q=\{\{q_{ji}(\omega )\nu (d\omega )\},\{V_{i}(\omega )\},\{\alpha _{i}\}\},
\label{2.15}
\end{equation}
consisting of\ three families of elements where:

\begin{itemize}
\item  $q_{ji}(\omega )\nu (d\omega ),i,j=1,...,N_{0};N_{0}\leq \infty $ are
complex scalar measures on a standard Borel space $(\Omega ,\mathcal{F)}$ ,
absolutely continuous with respect to a finite positive scalar measure $\nu
(\cdot ),$ with $q_{ii}(\omega )\geq 0$ and satisfying the orthonormality
relation 
\begin{equation}
\int_{\Omega }q_{ji}(\omega )\nu (d\omega )=\delta _{ji};  \label{2.16}
\end{equation}

\item  the $\alpha _{i},$ $i=1,...,N_{0}$ constitute a finite or countable
sequence of positive numbers that sum to $1;$

\item  each $V_{i}(\omega ),i=1,...,N_{0}$ is a$\ \nu $-measurable
operator-valued function with values being linear bounded operators on $%
\mathcal{H}$, satisfying the orthonormality relation 
\begin{equation}
\int_{\Omega }V_{j}^{\ast }(\omega )V_{i}(\omega )q_{ji}(\omega )\nu
(d\omega )=\delta _{ij}I  \label{2.17}
\end{equation}
and such that, for any $B\in \mathcal{F}$ and any index $i$, 
\begin{equation}
\int_{\omega \in B}V_{i}(\omega )q_{ii}(\omega )\nu (d\omega )\in \mathcal{B}%
(\mathcal{H)}.  \label{2.18}
\end{equation}
\end{itemize}

We let 
\begin{equation}
\nu _{i}(\mathrm{d}\omega )=q_{ii}(\omega )\nu (\mathrm{d}\omega ),
\label{2.19}
\end{equation}
\begin{equation}
\nu _{0}(d\omega )=\sum_{i}\alpha _{i}\nu _{i}(\mathrm{d}\omega )
\label{2.20}
\end{equation}
and refer to these as the \emph{input probability scalar measures}.

In case the index set for $i$ consists of one element only we drop the index
and assume that the probability density $q_{11}$ is identically $1$,
implying that $\nu (\cdot )$ is a probability measure, and we then say that
the QSR \ is $\emph{simple}$.

The $\nu $-measurable operator-valued functions $V_{i}(\omega )$, having the
properties (\ref{2.17}) and (\ref{2.18}) are called in [36] $quantum$\emph{\ 
}$stochastic$\emph{\ }$evolution$\emph{\ }$operators.$

Consider in general the statistical and stochastic description of a quantum
measurement, represented by a QSR.

The quantum instrument, corresponding to a direct quantum measurement, which
is determined by the quantum stochastic representation $Q,$ is given, for
all $B\in \mathcal{F}$ and all $Y\in \mathcal{B(H}),$ by 
\begin{equation}
\mathcal{N}(B)[Y]=\sum_{i}\alpha _{i}\mathcal{N}_{i}(B)[Y]  \label{2.21}
\end{equation}
with 
\begin{equation}
\mathcal{N}_{i}(B)[Y]=\int_{B}V_{i}^{\ast }(\omega )YV_{i}(\omega )\nu
_{i}(d\omega ).  \label{2.22}
\end{equation}
The probability scalar measure $\pi (\mathrm{d}\omega ;\rho _{0})$ on $%
\Omega ,$ defined by (\ref{2.4}), and the family of unnormalised posterior
states $\eta _{out}(\omega ;\rho _{0})$ on $\mathcal{H}$ are presented by
the following specifications 
\begin{equation}
\pi (\mathrm{d}\omega ;\rho _{0})=\sum_{i}\alpha _{i}\mathrm{tr}%
\{V_{i}(\omega )\rho _{0}V_{i}^{\ast }(\omega )\}\nu _{i}(\mathrm{d}\omega ),
\label{2.23}
\end{equation}
\begin{equation}
\eta _{out}(\omega ;\rho _{0})=\sum_{i}\alpha _{i}V_{i}(\omega )\rho
_{0}V_{i}^{\ast }(\omega )q_{ii}(\omega ).  \label{2.24}
\end{equation}
Introducing for every index $i=1,...,N_{0}$ the unnormalized posterior state 
\begin{equation}
\eta _{out}^{(i)}(\omega ;\rho _{0})=V_{i}(\omega )\rho _{0}V_{i}^{\ast
}(\omega ),  \label{2.25}
\end{equation}
we present the unnormalized posterior states (\ref{2.24}) and the
probability scalar measure (\ref{2.23}) as 
\begin{equation}
\eta _{out}(\omega ;\rho _{0})=\sum_{i}\alpha _{i}q_{ii}(\omega )\eta
_{out}^{(i)}(\omega ;\rho _{0})  \label{2.26}
\end{equation}
and 
\begin{equation}
\pi (\mathrm{d}\omega ;\rho _{0})=\sum_{i}\alpha _{i}\pi _{i}(\mathrm{d}%
\omega ;\rho _{0})  \label{2.27}
\end{equation}
with 
\begin{equation}
\pi _{i}(\mathrm{d}\omega ;\rho _{0})=\mathrm{\mathrm{tr}}\{\eta
_{out}^{(i)}(\omega ;\rho _{0})\}\nu _{i}(\mathrm{d}\omega ).  \label{2.28}
\end{equation}
The probability scalar measures $\pi _{i}(\cdot ;\rho _{0})$ and $\pi (\cdot
;\rho _{0})$ are called$\ output$ $probability$ $measures.$

Due to (\ref{2.8}), (\ref{2.26}) and (\ref{2.28}), for the associated
instrument $\mathcal{M}(\cdot )[\cdot ]$ we have the following
representation 
\begin{equation}
\mathcal{M}(B)[\rho _{0}]=\sum_{i}\alpha _{i}\mathcal{M}_{i}(B)[\rho _{0}],%
\text{ \ \ }\forall B\in \mathcal{F,}  \label{2.29}
\end{equation}
where 
\begin{equation}
\mathcal{M}_{i}(B)[\rho _{0}]=\int_{B}\eta _{out}^{(i)}(\omega ;\rho
_{0})\nu _{i}(\mathrm{d}\omega ),  \label{2.30}
\end{equation}
and, consequently, for any index $i$ the unnormalized posterior state $\eta
_{out}^{(i)}(\omega ;\rho _{0})$ can be considered as the Radon-Nikodym
derivative $\frac{\mathrm{d}\mathcal{M}_{i}}{\mathrm{d}\nu _{i}}$ of the $i$%
-th associated instrument $\mathcal{M}_{i}$ with respect to the input
probability measure $\nu _{i}.$

If the recorded result in the classical world is (only) that the outcome $%
\omega $ belongs to a certain set $B\in \mathcal{F}$ then the corresponding
probability of this and the ensuing knowledge of the (normalised) posterior
state of the quantum system are represented, respectively, as 
\begin{equation}
\pi (B;\rho _{0})=\int_{B}\pi (\mathrm{d}\omega ;\rho _{0})  \label{2.31}
\end{equation}
and 
\begin{equation}
\rho _{out}(B;\rho _{0})=\frac{\sum_{i}\alpha _{i}\int_{B}\eta
_{out}^{(i)}(\omega ;\rho _{0})\nu _{i}(\mathrm{d}\omega )}{\pi (B;\rho _{0})%
}.  \label{2.32}
\end{equation}

Due to the decompositions (\ref{2.22}), (\ref{2.26}) and (\ref{2.27}), in
the frame of the QSA $\mathcal{N}_{i}(\cdot )[\cdot ],$ $\mathcal{M}%
_{i}(\cdot )[\cdot ],$ $\eta _{out}^{(i)}(\omega ;\rho _{0})$, $\nu _{i}(%
\mathrm{d}\omega )$ and $\pi _{i}(\cdot ;\rho _{0})$ are interpreted to
present the instrument, the associated instrument, the unnormalized
posterior state, the input and the output probability distributions of
outcomes in the $i$-th random transition channel of a measurement,
respectively. The statistical weights of the different channels $i$\ are
given by $\alpha _{i}$, which are interpretable as probabilities.

Let the initial state of a quantum system be pure: $\rho _{0}=|\psi
_{0}\rangle \langle \psi _{0}|$. Due to the orthonormality relation (\ref
{2.17}) every pure state $V_{i}(\omega )\psi _{0},$ $i=1,...,N_{0}$, is
interpreted in the frame of the QSA as a \textit{posterior pure state outcome%
} in the Hilbert space $\mathcal{H}$ conditioned by the observed outcome $%
\omega $ and corresponding to the $i$-th random transition channel of the
quantum measurement.

For the observed outcome $\omega $ the probability of the posterior pure
state outcome $V_{i}(\omega )\psi _{0}$ in $\mathcal{H}$ is given by 
\begin{equation}
\theta _{i}(\omega )=\frac{\alpha _{i}q_{ii}(\omega )||V_{i}(\omega )\psi
_{0}||^{2}}{\sum_{j}\alpha _{j}q_{jj}(\omega )||V_{j}(\omega )\psi _{0}||^{2}%
}.  \label{2.33}
\end{equation}
The representation of the unconditional posterior state as 
\begin{equation}
\rho _{out}(\Omega ;\rho _{0})=\sum_{i}\alpha _{i}\int_{\Omega }V_{i}(\omega
)|\psi _{0}\rangle \langle \psi _{0}|V_{i}^{\ast }(\omega )\nu _{i}(\mathrm{d%
}\omega )  \label{2.34}
\end{equation}
is considered in the QSA as the usual statistical average over the posterior
pure state outcomes $|V_{i}(\omega )\psi _{0}\rangle \langle V_{i}(\omega
)\psi _{0}|,i=1,2,...$ with respect to the input probability distribution of
outcomes $\nu _{i}(\cdot )$ in channel $i$ and with respect to the different
channels, given with statistical weights $\alpha _{i},i=1,2,...$.

Physically, the concept of different random channels correponds, under the
same outcome $\omega ,$ to different underlying random quantum transitions
of the environment of a measuring device, which we can not, however, specify
with certainty.

Direct measurements, on a given quantum system, described by different QSR
are called \textit{stochastic representation equivalent} provided the QSR
give the same statistical and stochastic description,. For example, in the
frame of the QSA, the notion of a von Neumann (projective) measurement of a
discrete observable $Z=\sum_{j}\lambda _{j}P_{j\text{ }}$corresponds to the
stochastic representation equivalence class of direct measurements on $(%
\mathbb{R},\mathcal{B}(\mathbb{R}))$, for which the complete statistical and
stochastic description is determined by the von Neumann measurement
postulates [40], presented by the formulae (\ref{2.13}), (\ref{2.14}).

\section{Continuous-time direct measurements in the frame of QSA}

We would like now to introduce the general QSR describing a continuous, over
a time period $(0,T]$, direct quantum measurement. In this case the outcome $%
\omega $, characterizing continuous-time observation up to the moment $\
0<t\leq T$, is given by a record $\{x_{\tau }\}_{\tau \in (0,t]}$,
presenting a trajectory $x_{0}^{t}=\{x_{\tau }\}_{\tau \in (0,t]}$ in a
filtered standard Borel space $(\Omega ,\{\mathcal{F}_{t}\},\mathcal{F)}$,
and without essential loss of generality we think of $x_{t}$ as real-valued
and, for simplicity, we consider the case where the measure space $\Omega $
is represented by $D(0,T]$, the space of right continuous functions with
left limits, defined on $(0,T]$. In this case, for any time $t\in (0,T]$ the
trajectory $x_{0}^{t}$ is cadlag (continue a droite, limite a gauche).\
Further, $\mathcal{F}_{\tau }^{t}$ denotes the $\sigma $-algebra generated
by $x_{\tau }^{t}=\{x_{s}\}_{s\in (\tau ,t]}$ and we use the notation $%
\Omega _{\tau }^{t}$ for the restriction of $D(0,T]$ to the interval $(\tau
,t]$.

As discussed in Section 2, under the QSA for any generalized direct quantum
measurement there exists a unique QSR. Then, in particular, under a
continuous-time direct quantum measurement there must exist a unique QSR,
describing this special kind of a generalized direct measurement. The
elements of this QSR must have the time-wise properties that we now go on to
describe.

For simplicity, we consider only the case of simple QSRs.

Thus, in the frame of the QSA, for any continuous-time direct quantum
measurement, whose QSR is simple, at any moment of time $t\in (0,T]$ there
exist:

\begin{itemize}
\item  A unique input probability scalar measure $\nu _{0}^{t}(\cdot )$ on $%
(\Omega _{0}^{t}$,$\mathcal{F}_{t});$

\item  A unique family of measurable (with respect to $\mathcal{F}_{t}$)
operator-valued functions $\{V_{0}^{t}(x_{0}^{t}):x_{0}^{t}\in \Omega
_{0}^{t}\},$ defined $\nu _{0}^{t}$-almost everywhere on $\Omega _{0}^{t}$,
with values being bounded linear operators on $\mathcal{H}$ such that for
any $B_{0}^{t}\in \mathcal{F}_{t}$%
\begin{equation}
\int_{B_{0}^{t}}V_{0}^{t}(x_{0}^{t})\nu _{0}^{t}(\mathrm{d}x_{0}^{t})\in 
\mathcal{B(H)}\text{,}  \label{3.1}
\end{equation}
and the following normalisation relation is valid 
\begin{equation}
\int_{\Omega _{0}^{t}}(V_{0}^{t}(x_{0}^{t}))^{\ast }V_{0}^{t}(x_{0}^{t})\nu
_{0}^{t}(\mathrm{d}x_{0}^{t})=I.  \label{3.2}
\end{equation}
\end{itemize}

From (\ref{2.22}) it follows that for any continuous-time direct quantum
measurement with a simple QSR at any moment of time $t$ the instrument $%
\mathcal{N}_{0}^{t}(\cdot )[\cdot ]$ must be represented as: 
\begin{equation}
\mathcal{N}_{0}^{t}(B_{0}^{t})[Y]=\int_{B_{0}^{t}}(V_{0}^{t}(x_{0}^{t}))^{%
\ast }YV_{0}^{t}(x_{0}^{t})\nu _{0}^{t}(\mathrm{d}x_{0}^{t}),  \label{3.3}
\end{equation}
for $\forall B_{0}^{t}\in \mathcal{F}_{t},\forall Y\in \mathcal{B(H)},$ with
similar time-wise notation for the associated instrument $\mathcal{M}%
_{0}^{t}(\cdot )[\cdot ],$ the POV measure $M_{0}^{t}(\cdot )$, the output
laws $\pi _{0}^{t}(\cdot ;\rho _{0})$ and the family of unnormalized
posterior states $\{\eta _{out}^{t}(x_{0}^{t};\rho _{0}):$ $x_{0}^{t}\in
\Omega _{0}^{t}\},$ defined $\nu _{0}^{t}$-a.e. on $\Omega _{0}^{t}.$

Furthermore, we must include into the specification of the QSR, describing
the continuous-time direct measurement, the conditions that:

\begin{itemize}
\item  At all moments of time until $T$ the input probability scalar
measures (describing physically the measurement situation under which the
quantum system is not entangled with a measuring device) must be compatible
in time;

\item  The output laws $\pi _{0}^{t}(\cdot ;\rho _{0})$ should be also
compatible in time, corresponding to the compatibility in time of the POV
measures $M_{0}^{t}(\cdot )$;

\item  We assume that for any initial pure state $\psi _{0}\in \mathcal{H}$
under the continuous-time observation the posterior pure state outcome,
being a trajectory in the Hilbert space $\mathcal{H},$ presented at any
moment $t$ by the quantum stochastic evolution operator as $%
V_{0}^{t}(x_{0}^{t})\psi _{0}$, is continuous in $t$ from the right in the
norm on $\mathcal{H},$ $\nu _{0}^{t}$-a.e. on $\Omega _{0}^{t}$, with the
limit of $V_{0}^{t}(x_{0}^{t})\psi _{0}$ as $t\downarrow 0$ being equal to $%
\psi _{0}$. Under this specification, the situations where the quantum
system is isolated, are included into our representation as a special case.
In this case, for any $t$ the quantum stochastic operator $V_{0}^{t}$ does
not depend on the event $x_{0}^{t}$ in the classical world and is given by a
unitary operator $U(t,0)$, strongly continuous in $t$ for any $0<t\leq T$
both from the left and from the right.
\end{itemize}

Summing up all the above-mentioned points, we introduce the following
time-wise specification for the elements of the simple QSR, describing a
continuous-time direct quantum measurement:

\begin{itemize}
\item  A unique collection $\{\nu _{\tau }^{t}(\cdot ):0\leq \tau <t\leq T\}$
of input probability scalar measures such that every $\nu _{\tau }^{t}(\cdot
)$ on $(\Omega _{\tau }^{t},\mathcal{F}_{\tau }^{t})$ is the restriction of
the input probability scalar measure $\nu $: 
\begin{equation}
\nu _{\tau }^{t}(B_{\tau }^{t})=\nu (\Omega _{0}^{\tau }\times B_{\tau
}^{t}\times \Omega _{t}^{T});  \label{3.4}
\end{equation}

\item  A unique family $\{V_{0}^{t}(x_{0}^{t}):x_{0}^{t}\in \Omega
_{0}^{t},0<t\leq T\}$ of measurable (with respect to $\mathcal{F}_{t}$)
operator-valued functions $V_{0}^{t}(\cdot ):\Omega _{0}^{t}\rightarrow 
\mathcal{B(H)}$, defined $\nu _{0}^{t}$-almost everywhere on $\Omega
_{0}^{t},$ such that, for any $0<t\leq T$ and any $B_{0}^{t}\in \mathcal{F}%
_{t},$%
\begin{equation}
\int_{B_{0}^{t}}V_{0}^{t}(x_{0}^{t})\nu _{0}^{t}(\mathrm{d}x_{0}^{t})\in 
\mathcal{B(H)}\text{.}  \label{3.5}
\end{equation}
These operator-valued functions satisfy the normalisation relation 
\begin{equation}
\int_{\Omega _{0}^{t}}(V_{0}^{t}(x_{0}^{t}))^{\ast }V_{0}^{t}(x_{0}^{t})\nu
_{0}^{t}(\mathrm{d}x_{0}^{t})=I,  \label{3.6}
\end{equation}

and the initial condition 
\begin{equation}
\lim_{t\downarrow 0}||V_{0}^{t}(x_{0}^{t})\psi -\psi ||_{\mathcal{H}}=0,%
\text{ \ \ \ \ }\forall \psi \in \mathcal{H},  \label{3.7}
\end{equation}
$\nu _{0}^{t}$-a.e.on $\Omega _{0}^{t}$;

\item  A unique family $\{V_{\tau }^{t}(x_{0}^{t}):x_{0}^{t}\in \Omega
_{0}^{t},0<\tau \leq t\leq T\}$ of measurable (with respect to $\mathcal{F}%
_{t}$) operator-valued functions $V_{\tau }^{t}(\cdot ):\Omega
_{0}^{t}\rightarrow \mathcal{B}(\mathcal{H)},$ defined $\nu _{0}^{t}$-almost
everywhere on $\Omega _{0}^{t},$ and such that for any $0<\tau <t\leq T$ and
any $B_{\tau }^{t}\in \mathcal{F}_{\tau }^{t},$ $x_{0}^{\tau }\in \Omega
_{0}^{\tau }$%
\begin{equation}
\int_{B_{\tau }^{t}}V_{\tau }^{t}(x_{0}^{t})\nu _{\tau }^{t}(\mathrm{d}%
x_{\tau }^{t}|x_{0}^{\tau }))\in \mathcal{B(H)},  \label{3.8}
\end{equation}
and the following normalisation relation is valid\footnote{%
In (\ref{3.8}), (\ref{3.9}) $\nu _{\tau }^{t}(\mathrm{d}x_{\tau
}^{t}|x_{0}^{\tau })$ denotes the conditional probability measure on ($%
\Omega _{\tau }^{t},\mathcal{F}_{\tau }^{t})$.} 
\begin{equation}
\int_{\Omega _{\tau }^{t}}(V_{\tau }^{t}(x_{0}^{t}))^{\ast }V_{\tau
}^{t}(x_{0}^{t})\nu _{\tau }^{t}(\mathrm{d}x_{\tau }^{t}|x_{0}^{\tau })=I.
\label{3.9}
\end{equation}
These operator-valued functions are associated with the family of operators $%
\{V_{0}^{t}(x_{0}^{t})\}$ via the cocycle condition 
\begin{equation}
V_{\tau }^{t}(x_{0}^{t})=V_{s}^{t}(x_{0}^{t})V_{\tau }^{s}(x_{0}^{s}),
\label{3.10}
\end{equation}
valid for any $t\in (0,T],\tau \in \lbrack 0,T],s\in (0,T],\tau \leq s\leq t$%
, $\nu _{0}^{t}$-a.e. on $\Omega _{0}^{t}$ and where $V_{\tau
}^{t}(x_{0}^{t})|_{t=\tau }=I$. Furthermore, 
\begin{equation}
\lim_{t\downarrow \tau }||V_{\tau }^{t}(x_{0}^{t})\psi -\psi ||_{\mathcal{H}%
}=0,\text{ \ \ }\forall \psi \in \mathcal{H},  \label{3.11}
\end{equation}
$\nu _{0}^{t}$-a.e. on $\Omega _{0}^{t}.$
\end{itemize}

We shall show later that the cocycle relation (\ref{3.10}), together with
the normalisation relation (\ref{3.9}), ensures the compatibility of the
time dependent POV measures$.$

For the introduced time-dependent QSR we have the following collections of
time dependent quantum instruments 
\begin{equation}
\{\mathcal{N}_{0}^{t}(\cdot )[\cdot ]:0<t\leq T\},  \label{3.12}
\end{equation}
\begin{equation}
\mathcal{N}_{0}^{t}(B_{0}^{t})[Y]=\int_{B_{0}^{t}}(V_{0}^{t}(x_{0}^{t}))^{%
\ast }YV_{0}^{t}(x_{0}^{t})\nu _{0}^{t}(\mathrm{d}x_{0}^{t}),\text{ \ \ }%
\forall B_{0}^{t}\in \mathcal{F}_{t},\forall Y\in \mathcal{B(H)}
\label{3.13}
\end{equation}
and 
\begin{equation}
\{\mathcal{M}_{0}^{t}(\cdot )[\cdot ]):0<t\leq T\},  \label{3.14}
\end{equation}
\begin{equation}
\mathcal{M}_{0}^{t}(B_{0}^{t})[\kappa
]=\int_{B_{0}^{t}}V_{0}^{t}(x_{0}^{t})\kappa (V_{0}^{t}(x_{0}^{t}))^{\ast
}\nu _{0}^{t}(\mathrm{d}x_{0}^{t}),\ \ \ \forall B_{0}^{t}\in \mathcal{F}%
_{t},\forall \kappa \in \mathcal{T(H)}.  \label{3.15}
\end{equation}

The corresponding collection of time-dependent POV measures and the family
of time-dependent unnormalized posterior states are presented as 
\begin{equation}
\{M_{0}^{t}(\cdot ):0<t\leq T\},  \label{3.16}
\end{equation}
\begin{equation}
M_{0}^{t}(B_{0}^{t})=\int_{B_{0}^{t}}(V_{0}^{t}(x_{0}^{t}))^{\ast
}V_{0}^{t}(x_{0}^{t})\nu _{0}^{t}(\mathrm{d}x_{0}^{t}),\text{ \ \ }\forall
B_{0}^{t}\in \mathcal{F}_{t},  \label{3.17}
\end{equation}
and 
\begin{equation}
\{\eta _{out}^{t}(\cdot ;\rho _{0}):0<t\leq T\},  \label{3.18}
\end{equation}
\begin{equation}
\eta _{out}^{t}(x_{0}^{t};\rho _{0})=V_{0}^{t}(x_{0}^{t})\rho
_{0}(V_{0}^{t}(x_{0}^{t}))^{\ast },\text{ \ \ }\forall x_{0}^{t}\in \mathcal{%
\Omega }_{0}^{t},  \label{3.19}
\end{equation}
respectively.

The collection of time-dependent output laws has the form 
\begin{equation}
\{\pi _{0}^{t}(\cdot ;\rho _{0}):0<t\leq T\}  \label{3.20}
\end{equation}
with 
\begin{equation}
\pi _{0}^{t}(B_{0}^{t};\rho _{0})=\int_{B_{0}^{t}}\mathrm{tr}%
\{V_{0}^{t}(x_{0}^{t})\rho _{0}(V_{0}^{t}(x_{0}^{t}))^{\ast }\}\nu _{0}^{t}(%
\mathrm{d}x_{0}^{t}),\text{ \ \ }\forall B_{0}^{t}\in \mathcal{F}_{t}.
\label{3.21}
\end{equation}

At any moment of time $t$ and for any $B_{0}^{t}\in \mathcal{F}_{t}$ the
normalized posterior states are given by 
\begin{equation}
\rho ^{t}(B_{0}^{t};\rho _{0})=\frac{\int_{B_{0}^{t}}\eta
_{out}^{t}(x_{0}^{t};\rho _{0})\nu _{0}^{t}(\mathrm{d}x_{0}^{t})}{\pi
_{0}^{t}(B_{0}^{t};\rho _{0})}=\frac{\mathcal{M}_{0}^{t}(B_{0}^{t})[\rho
_{0}]}{\pi _{0}^{t}(B_{0}^{t};\rho _{0})}.  \label{3.22}
\end{equation}
In the sequel we shall also use the following notation for the unconditional
posterior state: 
\begin{equation}
\rho ^{t}(\rho _{0})\equiv \rho ^{t}(\Omega _{0}^{t};\rho _{0})=\mathcal{M}%
_{0}^{t}(\Omega _{0}^{t})[\rho _{0}]\text{,}  \label{3.23}
\end{equation}
satisfying the initial condition $\rho ^{t}(\rho _{0})\rightarrow \rho _{0}$
as $t\downarrow 0$ in the trace norm on $\mathcal{T(H)}$.

Due to the relations (\ref{3.4}), (\ref{3.9}) and (\ref{3.10}), for any $%
t>\tau ,$ we have the following martingale property 
\begin{equation}
\int_{\Omega _{\tau }^{t}}(V_{0}^{t}(x_{0}^{t}))^{\ast
}V_{0}^{t}(x_{0}^{t})\nu _{0}^{t}(\mathrm{d}x_{\tau }^{t}|x_{0}^{\tau
})=(V_{0}^{\tau }(x_{0}^{\tau }))^{\ast }V_{0}^{\tau }(x_{0}^{\tau }),
\label{3.24}
\end{equation}
from which it follows that the collection (\ref{3.16}) of time-dependent POV
measures is compatible in time, that is, for any $B_{0}^{\tau }\in \mathcal{F%
}_{\tau }$ we have: 
\begin{eqnarray}
M_{0}^{t}(B_{0}^{\tau }) &=&\int_{B_{0}^{\tau }}(V_{0}^{t}(x_{0}^{t}))^{\ast
}V_{0}^{t}(x_{0}^{t})\nu _{0}^{t}(\mathrm{d}x_{0}^{t})  \notag \\
&=&M_{0}^{\tau }(B_{0}^{\tau }).  \label{3.25}
\end{eqnarray}

For any $0<\tau <t$ and any $Y\in \mathcal{B(H)}$, for the instruments from
the collections (\ref{3.12}) and (\ref{3.14}) we have the following
properties: 
\begin{equation}
\mathcal{N}_{0}^{t}(\mathrm{d}x_{0}^{t})[Y]=\mathcal{N}_{0}^{\tau }(\mathrm{d%
}x_{0}^{\tau })[\mathcal{N}_{\tau }^{t}(\mathrm{d}x_{\tau }^{t}|x_{0}^{\tau
})[Y]],  \label{3.26}
\end{equation}
\begin{equation}
\mathcal{M}_{0}^{t}(\mathrm{d}x_{0}^{t})[\kappa ]=\mathcal{M}_{\tau }^{t}(%
\mathrm{d}x_{\tau }^{t}|x_{0}^{\tau })[\mathcal{M}_{0}^{\tau }(\mathrm{d}%
x_{0}^{\tau })[\kappa ]],  \label{3.27}
\end{equation}
where we have introduced the notation 
\begin{equation}
\mathcal{N}_{\tau }^{t}(\mathrm{d}x_{\tau }^{t}|x_{0}^{\tau })[Y]=(V_{\tau
}^{t}(x_{0}^{t}))^{\ast }YV_{\tau }^{t}(x_{0}^{t})\nu _{v}^{t}(\mathrm{d}%
x_{\tau }^{t}|x_{0}^{\tau }),\text{ \ \ \ \ \ \ \ }\forall Y\in \mathcal{B(H)%
}\text{,}  \label{3.28}
\end{equation}
\begin{equation}
\mathcal{M}_{\tau }^{t}(\mathrm{d}x_{\tau }^{t}|x_{0}^{\tau })[\kappa
]=V_{\tau }^{t}(x_{0}^{t})\kappa (V_{\tau }^{t}(x_{0}^{t}))^{\ast }\nu
_{\tau }^{t}(\mathrm{d}x_{\tau }^{t}|x_{0}^{\tau }),\text{ \ \ \ \ \ \ }%
\forall \kappa \in \mathcal{T(H)},  \label{3.29}
\end{equation}
for instruments $\mathcal{N}_{\tau }^{t}(\cdot |x_{0}^{\tau })[\cdot ]$ and $%
\mathcal{M}_{\tau }^{t}(\cdot |x_{0}^{\tau })[\cdot ],$ which we call
conditional.

Due to the properties (\ref{3.4})-(\ref{3.11}), the collection $\{\mathcal{M}%
_{0}^{t}(\Omega _{0}^{t})[\cdot ]:0<t\leq T\}$ with 
\begin{equation}
\mathcal{M}_{0}^{t}(\Omega _{0}^{t})[\kappa ]=\int_{\Omega
_{0}^{t}}V_{0}^{t}(x_{0}^{t})\kappa (V_{0}^{t}(x_{0}^{t}))^{\ast }\nu
_{0}^{t}(\mathrm{d}x_{0}^{t}),\ \ \forall \kappa \in \mathcal{T(H)},
\label{3.30}
\end{equation}
and the collection $\{\mathcal{M}_{\tau }^{t}(\Omega _{\tau
}^{t}|x_{0}^{\tau })[\cdot ]:0<\tau <t\leq T\}$ with 
\begin{equation}
\mathcal{M}_{\tau }^{t}(\Omega _{\tau }^{t}|x_{0}^{\tau })[\kappa
]=\int_{\Omega _{s}^{t}}V_{\tau }^{t}(x_{0}^{t})\kappa (V_{\tau
}^{t}(x_{0}^{t}))^{\ast }\nu _{\tau }^{t}(\mathrm{d}x_{\tau
}^{t}|x_{0}^{\tau }),\text{ \ \ \ }\forall \kappa \in \mathcal{T(H)},
\label{3.31}
\end{equation}
constitute families of time-dependent dynamical maps (cf., for example,
[29]).\ We shall call the dynamical map, which we introduce by (\ref{3.31}),
conditional\textit{.}

It follows also from (\ref{3.4}) - (\ref{3.11}) that for any $0<\tau <t\leq
T $ the time-dependent dynamical maps $\mathcal{M}_{0}^{t}(\Omega
_{0}^{t})[\cdot ]$, $\mathcal{M}_{\tau }^{t}(\Omega _{\tau }^{t}|x_{0}^{\tau
})[\cdot ]$ are strongly continuous in $t$ from the right with the following
limits: 
\begin{equation}
\lim_{t\downarrow 0}||\mathcal{M}_{0}^{t}(\Omega _{0}^{t})[\kappa ]-\kappa
||_{\mathcal{T(H)}}=0,  \label{3.32}
\end{equation}
\begin{equation}
\lim_{t\downarrow \tau }||\mathcal{M}_{\tau }^{t}(\Omega _{\tau
}^{t}|x_{0}^{\tau })[\kappa ]-\kappa ||_{\mathcal{T(H)}}=0,  \label{3.33}
\end{equation}
$\forall k\in \mathcal{T}(\mathcal{H})$, $\nu _{0}^{\tau }$-a.e. on $\Omega
_{0}^{\tau }$.

\section{Posterior pure state trajectories in a Hilbert space}

Together with an arbitrary pure initial state $\psi _{0}$, any given
collection of quantum stochastic evolution operators $\{V_{0}^{t}(\cdot
):0<t\leq T\}$, with the properties specified in Section 3, determines by 
\begin{equation}
\phi (t|x_{0}^{t})=V_{0}^{t}(x_{0}^{t})\psi _{0}  \label{4.1}
\end{equation}
a posterior pure state trajectory $\{\phi (\tau |x_{0}^{\tau })\}_{\tau \in
(0,T]}$ in the Hilbert space $\otimes _{\tau \in (0,T]}\mathcal{H}$
conditioned by the continuously observed trajectory $x_{0}^{T}$ in the
classical world.

Due to the specification of the time-dependent QSR, presented in (\ref{3.4})
- (\ref{3.11}) this trajectory is continuous in $t$ from the right for $%
\forall t\in (0,T]$ 
\begin{equation}
\lim_{t\downarrow \tau }||\phi (t|x_{0}^{t})-\phi (\tau |x_{0}^{\tau })||_{%
\mathcal{H}}=0.  \label{4.2}
\end{equation}
Furthermore, $\phi (t|x_{0}^{t})$ satisfies the limit condition 
\begin{equation}
\lim_{t\downarrow 0}\phi (t|x_{0}^{t})=\psi _{0}  \label{4.3}
\end{equation}
and, for any $0<t\leq T$, the following normalisation relation holds 
\begin{equation}
\int_{\Omega _{0}^{t}}||\phi (t|x_{0}^{t})||^{2}\nu (\mathrm{d}x_{0}^{t})=1.
\label{4.4}
\end{equation}

According the QSA, $\{\phi (\tau |x_{0}^{\tau })\}_{\tau \in (0,t]}$
presents a \textit{posterior pure state outcome under the continuous-time
measurement,} which depends on the observed trajectory $x_{0}^{t}$ in the
classical world.

Thus, for the case of measurement continuous\ in time until the moment $t$
both the observed outcome $x_{0}^{t}$\ in the classical world and the
posterior pure state outcome $\{\phi (\tau |x_{0}^{\tau })\}_{\tau \in
(0,t]} $\ in the Hilbert space $\otimes _{\tau \in (0,t]}\mathcal{H}$\ are
represented as trajectories.

Introduce also for any $\psi \in \mathcal{H}$ and any $s\leq t$ the notation 
\begin{equation}
\Phi (t,s;x_{0}^{t},\psi )=V_{s}^{t}(x_{0}^{t})\psi .  \label{4.5}
\end{equation}
Then from (\ref{3.9}) it follows that for any $\psi \in \mathcal{H}$ 
\begin{equation}
\int_{\Omega _{s}^{t}}||\Phi (t,s;x_{0}^{t},\psi )||^{2}\nu (\mathrm{d}%
x_{s}^{t}|x_{0}^{s})=||\psi ||^{2},  \label{4.6}
\end{equation}
and, due to the property (\ref{3.10}), we have the following relation 
\begin{equation}
\Phi (t;s;x_{0}^{t},\Phi (s,\tau ;x_{0}^{s},\psi ))=\Phi (t,\tau
;x_{0}^{t},\psi ),  \label{4.7}
\end{equation}
valid $\nu $-a.e. on $\Omega $ for any $t\in (0,T],\tau \in \lbrack
0,T],s\in (0,T],\tau \leq s\leq t.$ In particular, since $\phi
(s|x_{0}^{s})\equiv \Phi (s,0;x_{0}^{s},\psi _{0})$ we can also write 
\begin{equation}
\Phi (t;s;x_{0}^{t},\phi (s|x_{0}^{s}))=\phi (t|x_{0}^{t}).  \label{4.8}
\end{equation}

\bigskip

If the initial state $\rho _{0}$ of a quantum system is pure, that is $\rho
_{0}=|\psi _{0}\rangle \langle \psi _{0}|$, then under the continuous-time
direct measurement, described by the simple QSR, specified in Section 3, at
any moment $t\in (0,T]$ the probability (\ref{3.21}) of the observed record $%
x_{0}^{t}=\{x_{\tau }\}_{\tau \in (0,T]}$ to belong to a subset $%
B_{0}^{t}\subseteq \Omega _{0}^{t}$ is given by 
\begin{equation}
\pi _{0}^{t}(B_{0}^{t};\rho _{0})=\int_{B_{0}^{t}}||\phi
(t|x_{0}^{t})||^{2}\nu _{0}^{t}(dx_{0}^{t})  \label{4.9}
\end{equation}
and, due to (\ref{4.6}) and (\ref{4.8}), the collection $\{\pi
_{0}^{t}(\cdot ;\rho _{0}):t\in (0,T]\}$ of output laws is compatible in
time.

The conditional posterior state (\ref{3.21}) and the unconditional posterior
state (\ref{3.23}) are represented as 
\begin{equation}
\rho ^{t}(B_{0}^{t};\rho _{0})=\frac{\int_{B_{0}^{t}}|\phi
(t|x_{0}^{t})\rangle \langle \phi (t|x_{0}^{t})|\nu _{0}^{t}(dx_{0}^{t})}{%
\pi _{0}^{t}(B_{0}^{t};\rho _{0})},  \label{4.10}
\end{equation}
and 
\begin{equation}
\rho ^{t}(\rho _{0})=\int_{\Omega _{0}^{t}}|\phi (t|x_{0}^{t})\rangle
\langle \phi (t|x_{0}^{t})|\nu _{0}^{t}(dx_{0}^{t}),  \label{4.11}
\end{equation}
and, thus, correspond, respectively, to conditional and unconditional 
\textit{statistical averaging} over the posterior pure state outcomes $|\phi
(t|x_{0}^{t})\rangle \langle \phi (t|x_{0}^{t})|$ with respect to the input
probability distribution $\nu _{0}^{t}(\cdot )$.

\section{The case of Markov evolution}

Consider now the special case of continuous-time measurement under which the
quantum stochastic evolution operators $V_{\tau }^{t}(x_{0}^{t})$ satisfy
the following restriction 
\begin{equation}
V_{\tau }^{t}(x_{0}^{t})=V_{\tau }^{t}(x_{\tau }^{t}),\text{ \ \ for all }%
0\leq \tau <t,  \label{5.1}
\end{equation}
and these operators and the input probability measure $\nu (\cdot )$ are
such that, for any $0\leq \tau <t\leq T,$ the conditional instruments $%
\mathcal{M}_{\tau }^{t}(\Omega _{\tau }^{t}|x_{0}^{\tau })[\cdot ]$ and $%
\mathcal{N}_{\tau }^{t}(\Omega _{\tau }^{t}|x_{0}^{\tau })[\cdot ]$ become
unconditional in the sense that they do not depend on measurement outcomes
in the past, that is, $\nu _{0}^{\tau }$-a.e. on $\Omega _{0}^{\tau }$: 
\begin{eqnarray}
\mathcal{M}_{\tau }^{t}(\Omega _{\tau }^{t}|x_{0}^{\tau })[\cdot ]
&=&\int_{\Omega _{\tau }^{t}}V_{\tau }^{t}(x_{\tau }^{t})[\cdot ](V_{\tau
}^{t}(x_{\tau }^{t}))^{\ast }\nu _{\tau }^{t}(\mathrm{d}x_{\tau
}^{t}|x_{0}^{\tau })  \notag \\
&=&\int_{\Omega _{\tau }^{t}}V_{\tau }^{t}(x_{\tau }^{t})[\cdot ](V_{\tau
}^{t}(x_{\tau }^{t}))^{\ast }\nu _{\tau }^{t}(\mathrm{d}x_{\tau }^{t}):=%
\mathcal{M}_{\tau }^{t}(\Omega _{\tau }^{t})[\cdot ],  \label{5.2} \\
\mathcal{N}_{\tau }^{t}(\Omega _{\tau }^{t}|x_{0}^{\tau })[\cdot ]
&=&\int_{\Omega _{\tau }^{t}}(V_{\tau }^{t}(x_{\tau }^{t}))^{\ast }[\cdot
]V_{\tau }^{t}(x_{\tau }^{t})\nu _{\tau }^{t}(\mathrm{d}x_{\tau
}^{t}|x_{0}^{\tau })  \notag \\
&=&\int_{\Omega _{\tau }^{t}}(V_{\tau }^{t}(x_{\tau }^{t}))^{\ast }[\cdot
]V_{\tau }^{t}(x_{\tau }^{t})\nu _{\tau }^{t}(\mathrm{d}x_{\tau }^{t}):=%
\mathcal{N}_{\tau }^{t}(\Omega _{\tau }^{t})[\cdot ].  \label{5.3}
\end{eqnarray}

In this special case, due to the properties (\ref{3.26}) and (\ref{3.27}),
from (\ref{5.1}), (\ref{5.2}) and (\ref{5.3}) it follows that, for all $%
0<s<t\leq T,$%
\begin{equation}
\mathcal{N}_{0}^{t}(\Omega _{0}^{t})[\cdot ]=\mathcal{N}_{0}^{s}(\Omega
_{0}^{s}[\mathcal{N}_{s}^{t}(\Omega _{s}^{t})[\cdot ]],\text{ \ \ }\mathcal{M%
}_{0}^{t}(\Omega _{0}^{t})[\cdot ]=\mathcal{M}_{s}^{t}(\Omega _{s}^{t})[%
\mathcal{M}_{0}^{s}(\Omega _{0}^{s})[\cdot ]],  \label{5.4}
\end{equation}
and hence, the collection $\{\mathcal{M}_{s}^{t}(\Omega _{s}^{t})[\cdot
]:0\leq s<t\leq T\},$ where 
\begin{equation}
\mathcal{M}_{s}^{t}(\Omega _{s}^{t})[\cdot ]=\int_{\Omega
_{s}^{t}}V_{s}^{t}(x_{s}^{t})[\cdot ](V_{s}^{t}(x_{s}^{t}))^{\ast }\nu
_{s}^{t}(\mathrm{d}x_{s}^{t}),  \label{5.5}
\end{equation}
constitutes a family of time-dependent dynamical maps, satisfying the
cocycle relations (\ref{5.4}). Furthermore, the unconditional posterior
state $\rho ^{t}(\rho _{0})$, given, in general, by (\ref{3.23}), has here
the following\emph{\ Markov property} 
\begin{equation}
\rho ^{t}(\rho _{0})=\mathcal{M}_{s}^{t}(\Omega _{s}^{t})[\rho ^{s}(\rho
_{0})],\text{ \ \ }\forall 0\leq s<t.  \label{5.6}
\end{equation}
In (\ref{5.6}) we denote $\rho ^{0}(\rho _{0}):=\lim_{s\downarrow 0}\rho
^{s}(\rho _{0})=\rho _{0}$ where the limit is in the trace norm on $\mathcal{%
T}(\mathcal{H}).$

However, in contrast to the usual presentation of Markovian evolution of an
open system (cf.[29] and references cited therein) under the restrictions (%
\ref{5.1}) - (\ref{5.3}), the family of dynamical maps (\ref{5.5}) does not,
in general, represent a quantum dynamical semigroup.

Notice also that in the general QSA framework, considered in sections 3 and
4, the family (\ref{5.5}) does not generally satisfy the relations (\ref{5.4}%
), which are, however, usually assumed to be valid for the description of
continuous-time measurements in the frame of the operational approach
(cf.[16,29]), for example, in the case of the so-called instrumental
processes with independent increments [4-6, 29,30].

\bigskip

Given the condition (\ref{5.1}), let us, further, specify a sufficient
condition for the relations (\ref{5.2}), (\ref{5.3}) to be valid.

Suppose that, for all $0\leq s<t\leq T,$ the random operator $%
V_{s}^{t}(x_{s}^{t})$ is, under the law $\nu (\cdot ),$ stochastically
independent of $x_{0}^{s}.$ Then the relations (\ref{5.2}) and (\ref{5.3})
are fulfilled.

\bigskip

This sufficient condition is, in particular, true if the input probability
measure $\nu (\cdot )$ satisfies the relation: 
\begin{equation}
\nu (dx_{\tau _{1}}^{\tau _{2}}\times dx_{t_{1}}^{t_{2}})=\nu (dx_{\tau
_{1}}^{\tau _{2}})\nu (dx_{t_{1}}^{t_{2}}),  \label{5.7}
\end{equation}
for all $0\leq \tau _{1}<\tau _{2}\leq t_{1}<t_{2}\leq T$. In this simplest
case the conditional instruments (\ref{3.28}), (\ref{3.29}) become even
infinitesimally unconditional in the sense that, for any $0<s<t\leq T,$ the
instruments $\mathcal{N}_{s}^{t}(dx_{s}^{t}|x_{0}^{s})[\cdot ]$ and $%
\mathcal{M}_{s}^{t}(dx_{s}^{t}|x_{0}^{s})[\cdot ]$ do not depend on
measurement outcomes in the past: 
\begin{eqnarray}
\mathcal{N}_{s}^{t}(dx_{s}^{t}|x_{0}^{s})[\cdot ]
&=&(V_{s}^{t}(x_{s}^{t}))^{\ast }[\cdot ]V_{s}^{t}(x_{s}^{t})\nu _{s}^{t}(%
\mathrm{d}x_{s}^{t}):=\mathcal{N}_{s}^{t}(dx_{s}^{t})[\cdot ],  \notag \\
\mathcal{M}_{s}^{t}(dx_{s}^{t}|x_{0}^{s})[\cdot ]
&=&V_{s}^{t}(x_{s}^{t})[\cdot ](V_{s}^{t}(x_{s}^{t}))^{\ast }\nu _{s}^{t}(%
\mathrm{d}x_{s}^{t}):=\mathcal{M}_{s}^{t}(dx_{s}^{t})[\cdot ],  \label{5.8}
\end{eqnarray}
and hence, for any $B_{0}^{s}\in \mathcal{F}_{0}^{s},$ $B_{s}^{t}\in 
\mathcal{F}_{s}^{t}$%
\begin{equation}
\mathcal{N}_{0}^{t}(B_{0}^{s}\times B_{s}^{t})[\cdot ]=\mathcal{N}%
_{0}^{s}(B_{0}^{s})[\mathcal{N}_{s}^{t}(B_{s}^{t})[\cdot ]],\text{ \ \ }%
\mathcal{M}_{0}^{t}(B_{0}^{s}\times B_{s}^{t})[\cdot ]=\mathcal{M}%
_{s}^{t}(B_{s}^{t})[\mathcal{M}_{0}^{s}(B_{0}^{s})[\cdot ]].  \label{5.9}
\end{equation}
From (\ref{5.9}) the relations (\ref{5.2}), (\ref{5.3}) follow trivially.

\bigskip

Another possible situation where the sufficient condition is fulfilled
occurs when, under the law $\nu (\cdot ),$ $x_{t}$ is a process with
independent increments and the random operator $V_{s}^{t}(x_{s}^{t}),$ for
all $0\leq s<t\leq T$, depends stochastically only on the increments $%
\{(x_{r}-x_{u}):s\leq u<r\leq t\}$.

\bigskip

\textbf{Example. }In recent years the different stochastic calculus models
of continuous-time quantum measurement, based on the introduction of linear
(as well as non-linear) stochastic differential equations for a process $%
\{\psi _{t},t\in \lbrack 0,\infty )\}$ with values in a complex separable
Hilbert space $\mathcal{H}$,\ were intensively discussed in the mathematical
and physical literature.

As we have already mentioned in the introduction the type of stochastic
equation, used in all these presentations, corresponds to the quantum
filtering equation, derived in [7-10] for the quantum stochastic calculus
model of continuous-time indirect nondemolition measurements.

In the physical literature, in fact, only particular cases of such equations
were considered.

In the most general settings, the mathematical properties of this kind of
stochastic model on a filtered probability space $(\Omega ,\{\mathcal{F}%
_{t}\},\mathcal{F},P)$ were analysed in [2,4,6,27,29,30].

For the type of stochastic model in question it is postulated [6] that:

\begin{itemize}
\item  the (unnormalised) posterior state vector $\psi _{t}\in \mathcal{H}$
of the quantum system under continuous-time observation satisfies a
stochastic differential equation of Ito's type 
\begin{eqnarray}
\mathrm{d}\psi _{t} &=&-K_{t}\psi _{t-}\mathrm{d}t+\sum L_{kt}\psi _{t-}%
\mathrm{d}W_{kt}+  \label{5.10} \\
&&+\int_{\mathcal{Y}}(J_{t}\psi _{t-})(y)\overset{\sim }{\Pi }(\mathrm{d}y,%
\mathrm{d}t)  \notag
\end{eqnarray}
with a non-random initial condition $\psi _{0}=u\in \mathcal{H};$

\item  the $R^{d}$-valued observed output process is defined to be 
\begin{eqnarray}
X_{i}(t) &:&=\int_{0}^{t}c_{i}(s)\mathrm{d}s+\sum_{k=1}^{\infty
}\int_{0}^{t}a_{ik}(s)\mathrm{d}W_{ks}+  \label{5.11} \\
&&+\int_{\mathcal{Y}\times (0,t]}\varphi (g(y;s))g_{i}(y;s)\Pi (\mathrm{d}y,%
\mathrm{d}s)+  \notag \\
&&+\int_{\mathcal{Y}\times (0,t]}\frac{g_{i}(y;s)}{1+|g_{i}(y;s)|^{2}}%
\overset{\sim }{\Pi }(\mathrm{d}y,\mathrm{d}s),  \notag
\end{eqnarray}

with the functions 
\begin{eqnarray}
c &:&(0,\infty )\rightarrow R^{d},  \label{5.12} \\
a_{ik} &:&(0,\infty )\rightarrow R,  \notag \\
g &:&\mathcal{Y}\times (0,\infty )\rightarrow R^{d},  \notag \\
\varphi (z) &:&=\frac{|z|^{2}}{1+|z|^{2}},\text{ \ \ }z\in R^{d}  \notag
\end{eqnarray}

\item  $i=1,...,d;k=1,2,...$.
\end{itemize}

The following assumptions are supposed to hold [6] for the stochastic model,
defined by (\ref{5.10})-(\ref{5.12}):

\begin{itemize}
\item  For any $t\in (0,\infty )$ and any $k=1,2,...$ the operators $%
K_{t}\in \mathcal{B(H)}$, $L_{kt}\in \mathcal{B(H)}$, $J_{t}\in \mathcal{B}(%
\mathcal{H}$, $L^{2}(\mathcal{Y},\nu (\cdot );\mathcal{H}))$;

\item  $K_{t}+K_{t}^{\ast }=\sum_{k=1}^{\infty }L_{kt}^{\ast
}L_{kt}+J_{t}^{\ast }(I\otimes \gamma _{t})J_{t}$ with $\gamma _{t}$ being a
bounded multiplication operator on the space $L^{2}(\mathcal{Y},\nu (\cdot );%
\mathbb{C})$;

\item  The $W_{kt}$ are independent Brownian motions;

\item  $\Pi (\mathrm{d}y,\mathrm{d}t)$ is an adapted Poisson point process
on $\mathcal{Y}\times \lbrack 0,\infty )$ of intensity $\gamma _{t}(y)v(%
\mathrm{d}y)\mathrm{d}t$ and increments independent of the past;

\item  $\overset{\sim }{\Pi }(\mathrm{d}y,\mathrm{d}t)=\Pi (\mathrm{d}y,%
\mathrm{d}t)-\gamma _{t}(y)v(\mathrm{d}y)\mathrm{d}t$ is the compensated
Poisson process;

\item  $W=\{W_{kt}\}$ and $\Pi (\mathrm{d}y,\mathrm{d}t)$ are independent;
\end{itemize}

Due to these assumptions, under the law $P$ the output process (\ref{5.11})
is a process with independent increments. Let $\mathcal{E}_{s}^{t}$, 0$\leq
s\leq t$ denote the $\sigma $-algebra generated on $\Omega $ by $(X(r)-X(s))$%
, $r\in \lbrack s,t].$ Under a number of regularity conditions, it was
proved in [6] that:

\begin{itemize}
\item  The Cauchy problem for the equation (\ref{5.10}), with a nonrandom
initial condition $\psi _{0}=u$ at time $t=0$ has a unique (up to $P$%
-equivalence) solution $\psi _{t}=\Psi _{t}(0;\omega ;u)$ and for any $t$ $%
\geq 0$ the process $||\psi _{t}||^{2}$ is a positive martingale with 
\begin{equation}
\mathrm{E}_{P}\{||\psi _{t}||^{2}\}=||u||^{2};  \label{5.13}
\end{equation}

\item  For any $t\geq s$ the solution $\Psi _{t}(s;\omega ;\xi (\omega ))$
of the equation (\ref{5.10}) on the interval $(s,t]$ with the initial (at
time $s$) random condition $\xi (\omega ),$ where $\mathrm{E}_{P}\{||\xi
||^{2}\}<\infty ,$ satisfies the relation 
\begin{equation}
\mathrm{E}_{P}\{||\Psi _{t}(s;\cdot ;\xi )||^{2}\}=\mathrm{E}_{P}\{||\xi
||^{2}\}  \label{5.14}
\end{equation}
and, $P$-almost surely, 
\begin{equation}
\Psi _{t}(\tau ;\omega ;\Psi _{\tau }(s,\omega ;u))=\Psi _{t}(s;\omega ;u),%
\text{ \ \ }\forall t\geq \tau \geq s;  \label{5.15}
\end{equation}

\item  For any $u\in \mathcal{H}$, $A\in \mathcal{B(H)}$ and any $%
B_{s}^{t}\in \mathcal{E}_{s}^{t}$ the equation 
\begin{equation}
\mathrm{E}_{P}\{1_{B_{s}^{t}}\langle \Psi _{t}(s;\cdot ;u),A\Psi
_{t}(s;\cdot ;u)\rangle \}=\langle u,N_{s}^{t}(B_{s}^{t})[A]u\rangle ,
\label{5.16}
\end{equation}
defines a family of unconditional instruments $\{N_{s}^{t}(\Omega
_{s}^{t})[\cdot ]:0\leq s<t\}$ with the property (\ref{5.4}).
\end{itemize}

Recalling the results of Section 4, we see that the stochastic model of
continuous-time quantum measurement, based on the introduction of the
stochastic differential equation of the type (\ref{5.10}), corresponds to a
special case of our presentation, where for any $\omega \in \Omega _{s}^{t}$
the quantum stochastic evolution operator is defined by the equation: 
\begin{equation}
\Psi _{t}(s;\omega ;\psi )=V_{s}^{t}(\omega )\psi ,\text{ \ \ }\forall \psi
\in \mathcal{H}.  \label{5.17}
\end{equation}
The unconditional posterior state of a quantum system in this stochastic
model has the Markov property (\ref{5.6}).

\section{Measuring model of continuous-time direct quantum measurement}

For any moment of time $t\in (0,T]$ let us construct, up to unitary and
phase equivalence, a statistical realization, corresponding to the
time-dependent QSR, specified in Section 3. We shall refer to the resulting
realization as the measuring\ model of continuous-time direct quantum
measurement, corresponding to the (simple) QSR.

Let $\mathcal{H}(\nu )\equiv \mathcal{H(}\nu ,N;\Omega _{0}^{T})$ be the
direct integral [12], induced by a probability scalar measure $\nu (\cdot )$
and dimension function $N(x_{0}^{T})$ equal to identity $\nu $-a.e. on $%
\Omega _{0}^{T}$. For such a dimension function the direct integral $%
\mathcal{H}(\nu )$ is identical to $L_{2}(\Omega _{0}^{T},\nu ;\mathbb{C)}$.

The relation 
\begin{equation}
(X_{0}^{T}(B_{0}^{T})\varphi _{T})(x_{0}^{T})=\chi
_{B_{0}^{T}}(x_{0}^{T})\varphi _{T}(x_{0}^{T}),  \label{6.1}
\end{equation}
$\forall B_{0}^{T}\in \mathcal{F},$ \ $\forall \varphi _{T}\in \mathcal{H}%
(\nu ),$ holding $\nu $-a.e. on $\Omega _{0}^{T}$, defines a simple
projection-valued measure $X_{0}^{T}(\cdot ):\mathcal{F}\rightarrow \mathcal{%
B(\mathcal{H}(\nu ))}$ of the type $[X_{0}^{T}(\cdot )]=[\nu (\cdot )]$
(cf.[12,36]). Here $\chi _{B_{0}^{T}}$ denotes the indicator function of a
subset $B_{0}^{T}.$

Letting 
\begin{equation}
X_{\tau }^{t}(B_{\tau }^{t})=X_{0}^{T}(\Omega _{0}^{\tau }\times B_{\tau
}^{t}\times \Omega _{t}^{T}),  \label{6.2}
\end{equation}
the$\ $projection-valued measure $X_{0}^{T}(\cdot )$ defines by 
\begin{equation}
(X_{\tau }^{t}(B_{\tau }^{t})\varphi _{T})(x_{0}^{T})=\chi _{B_{\tau
}^{t}}(x_{\tau }^{t})\varphi _{T}(x_{0}^{T}),\ \text{\ }\forall B_{\tau
}^{t}\in \mathcal{F}_{\tau }^{t},\text{ \ }\forall \varphi _{T}\in \mathcal{%
\mathcal{H}}(\nu ),  \label{6.3}
\end{equation}
a collection $\{X_{\tau }^{t}(\cdot ):0\leq \tau <t\leq T\}$ of
time-dependent, mutually commuting and compatible, projection-valued
measures $X_{\tau }^{t}(\cdot )$ on the standard filtered Borel space $%
(\Omega _{0}^{T},\{\mathcal{F}_{t}\},\mathcal{F})$ with values in $\mathcal{%
B(\mathcal{H}}(\nu )\mathcal{)},$ satisfying for any $0\leq \tau <s<t\leq T$
the relation: 
\begin{equation}
X_{\tau }^{t}(B_{\tau }^{s}\times B_{s}^{t})=X_{s}^{t}(B_{s}^{t})X_{\tau
}^{s}(B_{\tau }^{s}),  \label{6.4}
\end{equation}
$\forall B_{s}^{t}\in \mathcal{F}_{s}^{t},\forall B_{\tau }^{s}\in \mathcal{F%
}_{\tau }^{s}$. In (\ref{6.3}) $x_{\tau }^{t}$ is the restriction of $%
x_{0}^{T}$ to the space $\Omega _{\tau }^{t}.$ For any $t>\tau \geq 0$ the
type $[X_{\tau }^{t}(\cdot )]$ equals $[\nu _{\tau }^{t}(\cdot )]$.

In the case considered, where $N(x_{0}^{T})=1$, $\nu $-a.e. on $\Omega
_{0}^{T}$, a base of measurability (cf.[12,36]) consists of only one element 
$e_{T}$, defined, up to unitary equivalence, by the relation $%
|e_{T}(x_{0}^{T})|=1$, $\nu $-a.e. on $\Omega _{0}^{T}.$ Since the measure $%
\nu (\cdot )$ is finite, $e_{T}\in \mathcal{\mathcal{H}}(\nu )$ and is an
element of maximum type for every projection-valued measure $X_{\tau
}^{t}(\cdot )$: 
\begin{equation}
\langle e_{T},X_{\tau }^{t}(\cdot )e_{T}\rangle _{\mathcal{H}(\nu )}=\nu
_{\tau }^{t}(\cdot ).  \label{6.5}
\end{equation}

Now, introduce the complex separable Hilbert space $\mathcal{K}(\nu )=%
\mathcal{H}\otimes \mathcal{\mathcal{H}(}\nu \mathcal{)}.$

Let $U_{\nu }(t,0)$ be a unitary operator on $\mathcal{K}(\nu )$, strongly
continuous in $t$ from the right for $\forall t\in (0,T]$, satisfying the
initial condition $s$-$\lim_{t\downarrow 0}U_{\nu }(t,0)=I$ (strong limit)
and such that for any vector $\psi \in \mathcal{H}$ the relation 
\begin{equation}
(U_{\nu }(t,0)(\psi \otimes e_{T}))(x_{0}^{T})=V_{0}^{t}(x_{0}^{t})\psi
\otimes e_{T}(x_{0}^{T}),  \label{6.6}
\end{equation}
is valid $\nu _{0}^{T}$-a.e. on $\Omega _{0}^{T}$. The unitary operator $%
U_{\nu }(t,0)$ is defined by the relation\ (\ref{6.6}) up to unitary
equivalence. The continuity conditions are required for the compatibility of
the properties of $U_{\nu }(t,0)$ with the properties of the quantum
stochastic evolution operators $V_{0}^{t}(x_{0}^{t}),$ specified by (\ref
{3.5})-(\ref{3.11}).

The statistical realization 
\begin{equation}
\{\mathcal{\mathcal{H}(}\nu \mathcal{)},|e_{T}\rangle \langle
e_{T}|,X_{0}^{t}(\cdot ),U_{\nu }(t,0)\}  \label{6.7}
\end{equation}
at any moment of time $t\in (0,T],$ presents on $\mathcal{\mathcal{H}(}\nu 
\mathcal{)}$ the invariant class $G(t)$ (cf.[36]) of unitarily and phase
equivalent separable statistical realizations, corresponding to the
time-dependent QSR, specified in Section 2.2.

For any $t\in (0,T]$ the representation of the instrument (\ref{3.13})
through the elements of the statistical realization (\ref{6.7}) is given by 
\begin{equation}
\mathcal{N}_{0}^{t}(B_{0}^{t})[Y]=\mathbf{E}_{|e_{T}\rangle \langle e_{T}|}%
\mathbf{[}U_{\nu }^{\ast }(t,0)(Y\otimes X_{0}^{t}(B_{0}^{t}))U_{\nu }(t,0)],
\label{6.8}
\end{equation}
where for any statistical operator $\sigma $ on $\mathcal{\mathcal{H}(}\nu 
\mathcal{)}$ the notation $\mathbf{E}_{\sigma }[\mathbf{\cdot ]}$ denotes
the normal completely positive bounded linear map $\mathbf{E}_{\sigma }%
\mathbf{[\cdot ]:}\mathcal{B}(\mathcal{K}(\nu ))\rightarrow \mathcal{B}(%
\mathcal{H})$, such that for $\forall Q\in \mathcal{B}(\mathcal{K}(\nu ))$
the relation 
\begin{equation}
\mathrm{tr}\{\rho \mathbf{E}_{\sigma }\mathbf{[}Q]\}=\mathrm{tr}\{(\rho
\otimes \sigma )Q\}  \label{6.9}
\end{equation}
is valid for any statistical operator $\rho $ on $\mathcal{H}$ [42].

The family (\ref{3.16}) of POV measures is represented by 
\begin{equation}
M_{0}^{t}(B_{0}^{t})=\mathbf{E}_{|e_{T}\rangle \langle e_{T}|}\mathbf{[}%
U_{\nu }^{\ast }(t,0)(I\otimes X_{0}^{t}(B_{0}^{t}))U_{\nu }(t,0)].
\label{6.10}
\end{equation}

Similar to (\ref{6.6}), introduce also for any $t\geq \tau >0$ the unitary
operator $U_{\nu }(t,\tau ),$ strongly continuous in $t$ from the right,
satisfying the relation $U_{\nu }(\tau ,\tau )=I$ and such that for $\forall
\psi \in \mathcal{H}$%
\begin{equation}
(U_{\nu }(t,\tau )(\psi \otimes e_{T}))(x_{0}^{T})=V_{\tau
}^{t}(x_{0}^{t})\psi \otimes e_{T}(x_{0}^{T})  \label{6.11}
\end{equation}
$\nu $-a.e. on $\Omega _{0}^{T}.$ Then we have the following relation for
the conditional instrument (\ref{3.28}): 
\begin{equation}
\int_{B_{0}^{t}}\mathcal{N}_{\tau }^{t}(dx_{\tau }^{t}|x_{0}^{\tau })[Y]\nu
_{0}^{\tau }(dx_{0}^{\tau })=\mathbf{E}_{|e_{T}\rangle \langle e_{T}|}%
\mathbf{[}U_{\nu }^{\ast }(t,\tau )(Y\otimes X_{0}^{t}(B_{0}^{t}))U_{\nu
}(t,\tau )],\text{ \ \ \ }\forall Y\in \mathcal{B(H})  \label{6.12}
\end{equation}
and, consequently, $\mathcal{N}_{\tau }^{t}(dx_{\tau }^{t}|x_{0}^{\tau
})[\cdot ]$ is the Radon-Nikodym\ derivative with respect to $\nu _{0}^{\tau
}(\cdot )$ of the instrument standing on the right hand side of (\ref{6.12}).

Due to (\ref{3.9}), (\ref{6.5}) and (\ref{6.12}), for any $t\geq \tau \geq
s>0$ we have 
\begin{eqnarray}
\nu _{0}^{s}(\cdot ) &=&\mathbf{E}_{|e_{T}\rangle \langle e_{T}|}[U_{\nu
}^{\ast }(t,\tau )(I\otimes X_{0}^{s}(\cdot ))U_{\nu }(t,\tau )]
\label{6.13} \\
&=&\mathbf{E}_{|e_{T}\rangle \langle e_{T}|}[I\otimes X_{0}^{s}(\cdot )] 
\notag
\end{eqnarray}
and, therefore, 
\begin{equation}
\mathbf{E}_{|e_{T}\rangle \langle e_{T}|}[U_{\nu }^{\ast }(t,\tau )[I\otimes
X_{0}^{s}(\cdot ),U_{\nu }(t,\tau )]]=0,  \label{6.14}
\end{equation}
where $T\geq t\geq \tau \geq s>0$. From (\ref{6.14}) it follows then that
the family 
\begin{equation}
\{U_{\nu }(t,\tau ):t\in (0,T];\tau \in \lbrack 0,T];t\geq \tau ;U_{\nu
}(\tau ,\tau )|_{\tau >0}=I;\text{ }s\text{-}\lim_{t\downarrow 0}U_{\nu
}(t,0)=I\}  \label{6.15}
\end{equation}
of unitary operators, strongly continuous in $t$ from the right, satisfying (%
\ref{6.6}) and (\ref{6.11}), has the property: 
\begin{equation}
\lbrack I\otimes X_{0}^{s}(\cdot ),U_{\nu }(t,\tau )](\psi \otimes e_{T})=0,
\label{6.16}
\end{equation}
$\forall \psi \in \mathcal{H};\forall t\geq \tau \geq s>0$.

Let $\mathcal{H}_{R}$ be a complex separable Hilbert space isometrically
isomorphic to $\mathcal{H}(\nu )$ by a unitary transform $R,$ that is, $%
\mathcal{\mathcal{H}}_{R}=RH\mathcal{(}\nu \mathcal{)}.$ The relation $%
P_{R}^{(\tau ,t]}(\cdot )=RX_{\tau }^{t}(\cdot )R^{-1}$ defines the family 
\begin{equation}
\{P_{R}^{(\tau ,t]}(\cdot ):T\geq t>\tau \geq 0\}  \label{6.17}
\end{equation}
of mutually commuting, compatible projection-valued measures $P_{R}^{(\tau
,t]}(\cdot ):\mathcal{F}_{\tau }^{t}\rightarrow \mathcal{B(H}_{R})$ of the
type $[\nu _{\tau }^{t}(\cdot )]$, satisfying 
\begin{equation}
P_{R}^{(\tau ,t]}(B_{s}^{t}\times B_{\tau }^{s})=P_{R}^{(\tau
,s]}(B_{s}^{t})P_{R}^{(s,t]}(B_{\tau }^{s}),  \label{6.18}
\end{equation}
$\forall B_{s}^{t}\in \mathcal{F}_{s}^{t},$ $\forall B_{\tau }^{s}\in 
\mathcal{F}_{\tau }^{s};$ $\forall t>s>\tau \geq 0$.

Let 
\begin{equation}
\{U_{R}(t,\tau ):t\in (0,T];\tau \in \lbrack 0,T];t\geq \tau ;U_{R}(\tau
,\tau )|_{\tau >0}=I;s\text{-}\lim_{t\downarrow 0}U_{R}(t,0)=I\}
\label{6.19}
\end{equation}
be the family of unitary operators on $\mathcal{K}_{R}=(I\otimes R)\mathcal{%
K(}\nu \mathcal{)}=\mathcal{H}\otimes \mathcal{H}_{R},$ corresponding to $%
U_{\nu }(t,\tau )$ on $\mathcal{K(}\nu \mathcal{)}$. Then 
\begin{equation}
U_{R}(t,\tau )=(I\otimes R)U_{\nu }(t,\tau )(I\otimes R^{-1}).  \label{6.20}
\end{equation}
Denote $f_{R}=Re_{T}.$ From (\ref{6.17}) and (\ref{6.20}) it follows that
for any $\psi \in \mathcal{H}$%
\begin{equation}
\mathbf{\lbrack }U_{R}(t,\tau ),I\otimes P_{R}^{(0,s]}(\cdot )](\psi \otimes
f_{R})=0,\text{ \ \ }\forall t\geq \tau \geq s>0.  \label{6.21}
\end{equation}
Furthermore, for any $t>\tau \geq 0$ and any $\psi \in \mathcal{H}$: 
\begin{eqnarray}
(I\otimes P_{R}^{(0,t]}(dx_{0}^{t}))U_{R}(t,\tau )(\psi \otimes f_{R})
&=&(V_{\tau }^{t}(x_{0}^{t})\otimes P_{R}^{(0,t]}(dx_{0}^{t}))(\psi \otimes
f_{R}),  \label{6.22} \\
U_{R}(t,\tau )(\psi \otimes f_{R}) &=&\int_{\Omega _{0}^{t}}(V_{\tau
}^{t}(x_{0}^{t})\otimes P_{R}^{(0,t]}(dx_{0}^{t}))(\psi \otimes f_{R}),
\label{6.23} \\
\langle f_{R},P_{R}^{(\tau ,t]}(\cdot )f_{R}\rangle _{\mathcal{H}_{R}}
&=&\nu _{\tau }^{t}(\cdot ),  \label{6.24}
\end{eqnarray}
where the relation (\ref{6.22}) should be understood in the infinitesimal
sense. For the description of the most general case of continuous-time
nondemolition measurement the relations (\ref{6.22})-(\ref{6.24}) were first
introduced in [35].

Due to (\ref{6.22})-(\ref{6.24}), we have 
\begin{eqnarray}
\mathbf{E}_{|f_{R}\rangle \langle f_{R}|}[U_{R}(t,0)] &=&\int_{\Omega
_{0}^{t}}V_{0}^{t}(x_{0}^{t})\nu _{0}^{t}(dx_{0}^{t}),  \label{6.25} \\
\mathbf{E}_{|f_{R}\rangle \langle f_{R}|}[(I\otimes
P_{R}^{(0,t]}(dx_{0}^{t}))U_{R}(t,\tau )] &=&V_{\tau }^{t}(x_{0}^{t})\nu
_{0}^{t}(dx_{0}^{t}).  \label{6.26}
\end{eqnarray}

From (\ref{3.10}), (\ref{6.21}) and (\ref{6.22}) it follows that for any $%
t\geq s\geq \tau >0$ and $\forall \psi \in \mathcal{H}$ the unitary
operators $U_{R}(t,\tau )$ (and, hence, also the unitary operators $U_{\nu
}(t,\tau )$) satisfy the following relation 
\begin{equation}
U_{R}(t,\tau )(\psi \otimes f_{R})=U_{R}(t,s)U_{R}(s,\tau )(\psi \otimes
f_{R}),  \label{6.27}
\end{equation}
which we call a cocycle property with respect to the vector $f_{R}\in 
\mathcal{H}_{R}.$

The statistical realization 
\begin{equation}
\{\mathcal{H}_{R},|f_{R}\rangle \langle f_{R}|,P_{R}^{(0,t]}(\cdot
),U_{R}(t,0)\}  \label{6.28}
\end{equation}
is unitarily equivalent to the statistical realization (\ref{6.7}) and at
any moment $t\in (0,T]$ presents, in general, the invariant class $G(t)$ of
unitarily and phase equivalent statistical realizations, corresponding to
the time-dependent QSR, specified in Section 3.

We shall call the 4-tuple 
\begin{equation}
\{\mathcal{H}_{R},|\varphi _{R}\rangle \langle \varphi
_{R}|,P_{R}^{(0,T]}(\cdot ),\{U_{R}(t,\tau ),:0\leq \tau <t\leq T\}\},
\label{6.29}
\end{equation}
represented by a simple projection-valued measure $P_{R}^{(0,T]}(\cdot
):\Omega _{0}^{T}\rightarrow \mathcal{B}(\mathcal{H}_{R})$ and a family of
unitary operators (\ref{6.19}), with properties (\ref{6.21}) and (\ref{6.22}%
) - (\ref{6.27}), respectively, the \textit{measuring\ model of
continuous-time direct quantum measurement, corresponding to a simple
time-dependent QSR.}

\section{Scheme for continuous-time indirect nondemolition measurement}

Building on [35], we now consider the scheme for continuous-time indirect
measurement presented in [7-10].\ This type of measurement implies that
indirect information about the quantum system $S$ is obtained via a direct
measurement upon another quantum system, say $R$ (with a Hilbert space $%
\mathcal{H}_{R})$, entangled with $S.$ The unitary evolution of the compound
system ($S$ plus $R$) on the complex separable Hilbert space $\mathcal{K}%
_{R}=\mathcal{H\otimes H}_{R}$ is described in the frame of the Hamiltonian
approach, while the description of a direct measurement upon the quantum
system $R$ from the point of view of QSA should be based on the introduction
of a corresponding QSR.

However, up to the present moment, the consideration in the physical and the
mathematical literature of continuous-time indirect observation on the
system $S$ has been given, in fact, only for a special case, where the POV
measure of the continuous-time direct measurement upon the quantum system $R$
is presented by the joint spectral measure of a family of self-adjoint,
time-dependent operators $\{Q_{H}(t):t\in (0,T]\}$ on $\mathcal{K}_{R}$,
mutually commuting 
\begin{equation}
\lbrack Q_{H}(t),Q_{H}(\tau )]=0,\text{ \ \ \ }\forall t,\tau \in (0,T]
\label{7.1}
\end{equation}
and corresponding in the Heisenberg picture to some observable of the
quantum system $R.$

A von Neumann observable $Q_{H}(t)$, $t\in (0,T]$, satisfying the condition (%
\ref{7.1}) is usually termed nondemolition [29,44,35] or self-nondemolition
(cf. [7-10] and references there).

However, as was pointed out in [7-10], the condition (\ref{7.1}) alone does
not ensure the existence, at any moment of time $t\in (0,T],$ of an
instrument (with respect to the quantum system $S)$ which describes, via (%
\ref{2.3}), conditional expectations of any von Neumann $S$-system
observable $Z$ under continuous-time indirect measurement and, consequently,
allows to introduce the family of posterior states (cf.(\ref{2.8})).

That is why, in [7-10], along with the condition (\ref{7.1}) there was also
introduced an additional condition, specified below by (\ref{7.2}). These
two conditions are required to represent \textit{continuous-time indirect
nondemolition measurement} and are announced in [7-10] as ``\textit{%
principles of continuous in time nondemolition observation''}. \bigskip

Let $\{U(t,\tau ):t,\tau \in \lbrack 0,T]\}$ be the cocycle of unitary
operators, describing the evolution of the compound system ($S$ plus $R$) in
the interaction picture, induced by the free dynamics of the system $R$
(cf., for example, [35]). Then, according to the definition given in [7-10],
under continuous-time indirect nondemolition measurement:

\begin{itemize}
\item  there must exist the nondemolition observable $Q_{H}(t)=U^{\ast
}(t,0)(I\otimes Q_{R}(t))U(t,0)$, corresponding to some free dynamical
observable $Q_{R}(t)$ of the system $R$;

\item  at any moment of time $t\in \lbrack 0,T]$ any von Neumann $S$-system
observable $Z_{H}(t)=U^{\ast }(t,0)(Z\otimes I)U(t,0),$ $Z\in \mathcal{B(H)}$%
, where $Z=Z^{\ast }$, must commute with$\ $the observables $Q_{H}(s)$ at
all previous moments of time: 
\begin{equation}
\lbrack Z_{H}(t),Q_{H}(s)]=0,\text{ \ \ }\forall t\geq s\geq 0.  \label{7.2}
\end{equation}
\end{itemize}

Suppose, for simplicity, that for the family of self-adjoint, mutually
commuting operators $\{Q_{H}(t),$ $t\in (0,T]\}$ its joint spectrum [12]
coincides with $\Omega _{0}^{T}=D(0,T].$

We are now in position to prove the following statement.

\bigskip

\textbf{Proposition.}\textit{\ In the most general case, that is without
specifying a concrete nondemolition measurement model, the simultaneous
fulfilment of conditions }(\ref{7.1})\textit{\ and }(\ref{7.2})\textit{\ is
equivalent to the following:}

\begin{itemize}
\item  \textit{the family of self-adjoint operators} $\{Q_{R}(t):t\in
(0,T]\} $\textit{\ is a family of mutually commuting operators such that
their joint spectral projection-valued measure }$P_{R}^{(0,T]}(\cdot
):\Omega _{0}^{T}\rightarrow \mathcal{B(H}_{R}\mathcal{)}$, \textit{for any }%
$T\geq t\geq \tau >0,$ \textit{satisfies the commutativity relation} 
\begin{equation}
\lbrack U(t,\tau ),I\otimes P_{R}^{(0,\tau ]}(\cdot )]=0,  \label{7.3}
\end{equation}
\textit{where} $P_{R}^{(0,t]}(B_{0}^{t})=P_{R}^{(0,T]}(B_{0}^{t}\times
\Omega _{t}^{T})$ for $\forall B_{0}^{t}\in \mathcal{F}_{t}.$ (\textit{Note
that } 
\begin{equation}
\lbrack U(t,\tau ),I\otimes Q_{R}(s)]=0,  \label{7.4}
\end{equation}
$\forall $ $t\geq \tau \geq s>0$, \textit{presents an equivalent formulation
of the relation} (\ref{7.3})).\bigskip
\end{itemize}

\textbf{Proof}. Let the (\ref{7.1}) and (\ref{7.2}) be satisfied. Then from
the condition (\ref{7.1}) it follows (cf.[12]) that there exists a joint
projection-valued measure $P_{H}^{(0,T]}(\cdot ):\Omega _{0}^{T}\rightarrow 
\mathcal{B}\mathcal{(H\otimes H}_{R})$ such that for any $t\in (0,T]$%
\begin{equation}
Q_{H}(t)=\int_{\Omega _{0}^{T}}x_{t}P_{H}^{(0,T]}(dx_{0}^{T})=\int_{\Omega
_{0}^{t}}x_{t}P_{H}^{(0,t]}(dx_{0}^{t}),  \label{7.5}
\end{equation}
where the projection-valued measure $%
P_{H}^{(0,t]}(B_{0}^{t})=P_{H}^{(0,T]}(B_{0}^{t}\times \Omega _{t}^{T}).$

For any $Z\in \mathcal{B(H)}$, $Z=Z^{\ast }$, the commutativity relation (%
\ref{7.2}) is then equivalent to 
\begin{equation}
\lbrack Z_{H}(t),P_{H}^{(0,t]}(\cdot )]=0  \label{7.6}
\end{equation}
and, hence, to 
\begin{equation}
\lbrack Z\otimes I,U(t,0)P_{H}^{(0,t]}(\cdot )U^{\ast }(t,0)]=0.  \label{7.7}
\end{equation}

Since (\ref{7.7}) is valid for any von Neumann $S$-system observable $Z\in 
\mathcal{B(H)}$, by the commutation theorem of von Neumann algebras the
projection-valued measure $U(t,0)P_{H}^{(0,t]}(\cdot )U^{\ast }(t,0)$ must
have the form: 
\begin{equation}
U(t,0)P_{H}^{(0,t]}(\cdot )U^{\ast }(t,0)=I\otimes P_{R}^{(0,t]}(\cdot ).
\label{7.8}
\end{equation}
From (\ref{7.8}) it follows that the relations 
\begin{eqnarray}
P_{H}^{(0,\tau ]}(\cdot ) &=&U^{\ast }(\tau ,0)(I\otimes P_{R}^{(0,\tau
]}(\cdot ))U(\tau ,0)  \label{7.9} \\
&=&U^{\ast }(t,0)(I\otimes P_{R}^{(0,\tau ]}(\cdot ))U(t.0)  \notag
\end{eqnarray}
are valid for any $t\geq \tau >0$. The relation (\ref{7.3}) follows from (%
\ref{7.9}) trivially.

Furthermore, due to $Q_{H}(t)=U^{\ast }(t,0)(I\otimes Q_{R}(t))U(t,0)$ and
the relations (\ref{7.8}) and (\ref{7.9}), for any $t>0$ we have the
following representation: 
\begin{equation}
Q_{R}(t)=\int_{\Omega _{0}^{T}}x_{t}P_{R}^{(0,T]}(dx_{0}^{T})=\int_{\Omega
_{0}^{t}}x_{t}P_{R}^{(0,t]}(dx_{0}^{t}).  \label{7.10}
\end{equation}
Consequently, the von Neumann observable $Q_{R}(t),$ $t\in (0,T]$ is also
nondemolition.

The proof of the converse statement is straightforward.

\bigskip

We would like to emphasize here that although the conditions (\ref{7.1}) and
(\ref{7.2}) do not imply any concrete measurement model, the consideration
of continuous-time indirect nondemolition measurement (cf.[7-10] and
references therein), leading to the derivation of the quantum filtering
equation, was presented only in the frame of quantum stochastic calculus.
The measurement model of quantum stochastic calculus is essentially
Markovian. That is why, as already pointed out in the Introduction and
Section 5, under the scheme of continuous-time indirect nondemolition
measurement, the quantum filtering equation, introduced in [7-10], as well
as its further analogues [4-6, 27,29], correspond to quite special
stochastic models, which are Markovian.

\section{The scheme for continuous-time indirect nondemolition measurement
as a special measuring model of continuous-time measurement}

In this Section we show that :

\begin{itemize}
\item  For the general measuring model (\ref{6.29}) of continuous-time
direct quantum measurement there exists a uniquely determined family $%
\{Q_{R}(t),t\in (0,T]\}$ of mutually commuting (and hence nondemolition)
self-adjoint operators on $\mathcal{H}$, defined on a common domain $D.$ For
any $0<\tau \leq t\leq T$ the joint spectral measure of these operators
satisfies the relation 
\begin{equation}
\lbrack U(t,\tau ),I\otimes P_{R}^{(0,\tau ]}(\cdot )](\psi \otimes f_{R})=0.
\label{8.1}
\end{equation}

Since the condition (\ref{7.3}) is only sufficient for the relation (\ref
{8.1}) to be valid and since, in contrast to the model of ''continuous-time
indirect nondemolition measurement'', the unitary operators $U(t,\tau
),0\leq \tau <t\leq T$ in (\ref{6.29}) are strongly continuous in $t$ only
from the right, the model of continuous-time `indirect nondemolition
measurement' represents only a special case of the measuring model (\ref
{6.29}) of continuous-time observation of a quantum system.

\item  For the most general (that is, not only in the frame of quantum
stochastic calculus) model of continuous-time nondemolition measurement with
initial state of the system $R$ being pure, under some further technical
(for simplicity) restrictions, specified below, there exists the uniquely
defined simple time-dependent QSR, introduced in Section 2 and satisfying
properties (\ref{3.4}) - (\ref{3.11}).\bigskip
\end{itemize}

Consider the first point.

Let $P_{R}^{(0,T]}$ be the projection-valued measure of a measuring model (%
\ref{6.29}). Introduce the system of self-adjoint operators $\{Q_{R}(t):t\in
(0,T]\}$ given by 
\begin{equation}
Q_{R}(t)=\int_{\Omega _{0}^{T}}x_{t}P_{R}^{(0,T]}(dx_{0}^{T})=\int_{\Omega
_{0}^{t}}x_{t}P_{R}^{(0,t]}(dx_{0}^{t}).  \label{8.2}
\end{equation}
These operators are mutually commuting (cf.[12]) with a common domain 
\begin{equation}
D=\{f\in \mathcal{H}_{R}:\int_{\Omega _{0}^{T}}(x_{t})^{2}\nu
_{f}(dx_{0}^{T})<\infty ,\forall t\in (0,T]\},  \label{8.3}
\end{equation}
where the probability scalar measure $\nu _{f}(dx_{0}^{T})=\langle
f,P_{R}^{(0,T]}(dx_{0}^{T})f\rangle $ on $\Omega _{0}^{T}.$ The relation (%
\ref{8.1}) corresponds then to (\ref{6.21}).

\bigskip

Let us now prove the second point.

Come back to the notation of Section 7. Let $f_{R}$ be the initial state of
the quantum system $R$ and let the conditions (\ref{7.1}) and (\ref{7.2}) be
satisfied. Suppose also, for simplicity, that the joint spectrum of the
family of nondemolition observables $\{Q_{H}(t):t\in (0,T]\}$ coincides with 
$\Omega _{0}^{T}=D(0,T].$

Then, according to the consideration in Section 7, $\{Q_{R}(t):t\in (0,T]\}$
must be a family of self-adjoint mutually commuting observables with the
joint projection-valued measure $P_{R}^{(0,T]}(\cdot ):\Omega
_{0}^{T}\rightarrow \mathcal{B(H}_{R}\mathcal{)},$ satisfying the relation (%
\ref{7.3}).

The relation 
\begin{equation}
\langle f_{R,}P_{R}^{(0,T]}(\cdot )f_{R}\rangle =\nu _{0}^{T}(\cdot )
\label{8.6}
\end{equation}
determines the probability scalar measure $\nu _{0}^{T}(\cdot )$ on the
filtered space $(\Omega _{0}^{T},\{\mathcal{F}_{t}\},\mathcal{F})$. For
simplicity, suppose that $P_{R}^{(0,T]}(\cdot )$ is simple. The family of
quantum stochastic evolution operators $V_{\tau }^{t}(\cdot ):\Omega
_{0}^{t}\rightarrow \mathcal{B}(\mathcal{H)},$ $\ \forall t,\tau \in (0,T]\}$%
, is then introduced\ similarly to (\ref{6.11}) (cf. also [36]) and has the
properties (\ref{3.5}) - (\ref{3.11}).

\section{Concluding remarks}

A full description and classification, in terms of invariants, of all
possible integral representations of any given quantum instrument were
presented in [36] where the interpretation of the derived mathematical
results for the quantum measurement theory was also proposed and discussed.
In the present paper we consider the further development of the general
quantum stochastic approach, introduced in [36,37], for the description, in
the most general case, of statistical and stochastic aspects under
continuous-time measurement.

Specifying, in general, the time-wise properties of a quantum stochastic
evolution operator describing the stochastic evolution of an open quantum
system subjected to continuous-time observation, we introduce the notion of
a conditional quantum instrument and discuss the properties of the families
of time-dependent quantum instruments that describe a continuous-time
measurement. We present also the time-wise specifications of compatible in
time outcome laws and the formulae for the conditional and unconditional
posterior states.

Further, we define, in the most general case and without assuming any Markov
property, the notion of the posterior pure state trajectories in a Hilbert
space and present their (compatible in time) probabilistic description. The
restrictions, under which the stochastic evolution of a continuously
observed quantum system is Markovian, are also specified.

We construct a\ `canonical` measuring model of a continuous-time observation
of an open quantum system and prove that, formally, the scheme for
continuous-time indirect nondemolition measurement represents a special case
of this model.

\bigskip

\begin{center}
\textbf{Acknowledgement}
\end{center}

\begin{quote}
The present work was carried out in the framework of MaPhySto (Centre for
Mathematical Physics and Stochastics), which is funded by the Danish
National Research Foundation.
\end{quote}

\end{document}